\documentclass[11pt, twoside, a4paper]{article}
\usepackage{feynmp}
\usepackage{amsmath, amsthm, amssymb, array, dsfont, mathtools}
\usepackage[margin=1.1in]{geometry}
\usepackage[pdftex]{graphicx}
\usepackage{bm, comment, tikz, tikz-cd, setspace, mathrsfs, microtype, siunitx, booktabs, cancel, caption,
colortbl, csquotes, grffile, multirow, listings, pgfplots, xcolor, ulem}
\usepackage{subfig}
\usepackage{authblk}
\usepackage{cite}

\allowdisplaybreaks
\pgfplotsset{compat=1.12}

\definecolor{MyDarkBlue}{rgb}{0.15,0.15,0.45}
\usepackage[linktocpage=true]{hyperref}
\hypersetup{
colorlinks=false,
citecolor=MyDarkBlue, 
linkcolor=MyDarkBlue,
urlcolor=MyDarkBlue,
}

\makeatletter
\renewcommand*{\@textcolor}[3]{%
  \protect\leavevmode
  \begingroup
    \color#1{#2}#3%
  \endgroup
}
\makeatother

\newcommand{\del}{\partial}

\newcommand{\bt}[1]{\textbf{#1}}

\newcommand{\bs}[1]{\boldsymbol{#1}}
\newcommand{\wt}{\widetilde}
\newcommand{\hbt}[1]{\hat{\textbf{#1}}}

\renewcommand{\b}{\beta}
\newcommand{\g}{\gamma}

\newcommand{\e}{\epsilon}

\newcommand{\h}{\eta}
\renewcommand{\th}{\theta}

\newcommand{\m}{\mu}
\newcommand{\n}{\nu}

\newcommand{\p}{\pi}
\renewcommand{\r}{\rho}
\newcommand{\s}{\sigma}

\newcommand{\f}{\phi}
\newcommand{\ch}{\chi}

\newcommand{\D}{\Delta}

\renewcommand{\S}{\Sigma}

\renewcommand{\O}{\Omega}

\newcommand{\ol}{\overline}

\newcommand{\be}{\begin{equation}}
\newcommand{\ee}{\end{equation}}

\numberwithin{equation}{section}

\begin{document}
\unitlength = 1mm
\setlength{\parskip}{1em}

\title{\bt{Charging up Boosted Black Holes}}
\author[1]{Prakruth Adari}
\author[2]{Roman Berens}
\author[3]{Janna Levin}
\affil[1]{\fontsize{9}{12}\selectfont \textit{Department of Physics and Astronomy, Stony Brook University, Stony Brook, NY 11794}}
\affil[2]{\fontsize{9}{12}\selectfont \textit{Department of Physics, Columbia University, New York, NY 10027}}
\affil[3]{\fontsize{9}{12}\selectfont \textit{Department of Physics and Astronomy, Barnard College of Columbia University, New York, NY 10027}}

\date{}
\maketitle

\begin{center}
\fontsize{10}{12}\selectfont
Email: \href{mailto:prakruth.adari@stonybrook.edu}{prakruth.adari@stonybrook.edu},
\href{mailto:roman.berens@columbia.edu}{roman.berens@columbia.edu},
\href{mailto:janna@astro.columbia.edu}{janna@astro.columbia.edu}
\end{center}

\begin{abstract}

Contrary to a prevailing assumption that black holes would swiftly discharge, we argue that black holes can charge preferentially when boosted through an ambient magnetic field. Though the details are very different, the preference for charge is related to the precipitation of the Wald charge on a spinning black hole in an ambient magnetic field. The gravito-electrodynamics upstage naive arguments about screening electric fields--in vacuum--in determining the value of the charge accrued. Charged test particles, which build up the black hole charge, exhibit chaotic behavior as evidenced by fractal basin boundaries between dynamical regions. Charged, boosted black holes will generate their own electromagnetic fields and thereby their own luminous signatures, even if they are initially bare. We therefore add boosted black holes to the growing list of potentially observable black hole signatures, alongside black hole batteries and black hole pulsars. The implications should be relevant for supermassive black holes that are boosted relative to a galactic magnetic field as well as black holes merging with magnetized neutron stars.  

\end{abstract}

\newpage
\tableofcontents

\newpage
\section{Introduction}

Detailed direct observations of black holes this century motivate a renewed interest in the electromagnetic potentiality of black holes. 
Even bare, initially dark black holes cleverly exploit ambient magnetic fields---whether seeded by cohabitating neutron stars or anchored in a diffuse background---to power novel luminous channels. For instance, black hole batteries turn on during merger with a highly-magnetized neutron star, leading to faint electromagnetic counterparts to gravitational wave observations \cite{McWilliams:2011zi, DOrazio:2013ngp, Mingarelli:2015bpo, DOrazio:2015jcb}. Relatedly, a spinning black hole in an external magnetic field will preferentially acquire charge, as shown in 
an oft overlooked observation attributable to Wald \cite{Wald:1974np}. A spinning charge generates a magnetic dipole inducing a black hole pulsar \cite{Levin:2018mzg}. Even the Penrose process is enormously amplified around charged and magnetized black holes \cite{Gupta:2021vww}. All of these phenomena---black hole batteries, black hole pulsars, the electromagnetic Penrose process---are distinct yet interrelated. They are each consequences of the motion of a black hole through a magnetic field. 

We extend the range of luminous phenomena by considering black holes that are relativistically boosted through a uniform magnetic field.
Like spinning black holes, these boosted black holes will acquire charge, although the details are rather different. The boosted black hole perceives both a magnetic field $B_0$ and an electric field $E_0 = \b B_0$ due to the Lorentz transformation of the ambient magnetic field $\g^{-1} B_0$. Nearby charges will respond to the boosted black hole resulting in a charged boosted black hole, as we show. 

The gravito-electrodynamics are far more complicated for a boosted black hole than a spinning black hole. The boost breaks axial symmetry so a constant of geodesic motion is lost, namely the angular momentum. The Carter constant is also lost \cite{Igata:2010ny, Kolar:2015cha, Levin:2018mzg}. The only generic constants of motion retained for particles around the black hole are the energy and the timelike constraint on the four-velocity. With the loss of constants of motion, dynamics in phase space are no longer tightly constrained and the window to chaotic dynamics is opened. We therefore anticipate, and observe, extremely intricate orbits, reflecting the underlying chaos.

The complicated dynamics also highlight another general conclusion: dynamics upstage screening. By screening, we mean the condition that the electric potential difference between the horizon and infinity is zero in the vacuum case. By this we mean that while charges may be inclined to distribute themselves so as to screen an electric field, the actual dynamics may not comply with this inclination. Consequently, the charge acquired depends sensitively on the initial data and distribution of the test charges, as we show. Some distributions lead to a charged black hole and others do not. Qualitatively, the magnitude of the black hole charge will be set by the boost $\b$, the ambient magnetic field $B_0$, the mass of the black hole $M$. Based on units, this would indicate, again only qualitatively, $Q \propto \b B_0 M^2$. Due to the extreme sensitivity to initial conditions of chaotic dynamics, there is no simple quantitative formula to quote for the anticipated magnitude of the charge. Completely spatially symmetric distributions, lead to an average of zero black hole charge, although test charges continue to flux along the field lines. (Note that by ``symmetric'' here we simply mean spatially symmetric distribution of charged particles about the axis of the boost.) Spatially asymmetric distributions of ambient charges will lead to charged boosted black holes. 

Although an assessment of realistic luminosities is beyond the scope of this work, it is important to note that a boosted charge creates its own magnetic field, which in turn leads to its own electromagnetic signatures \cite{Chen:2021sya}. Ultimately, a charged boosted black hole could have observable implications for supermassive black holes like M87* or Sagittarius A* as well as black hole/neutron star mergers detectable by the LIGO-VIRGO-KAGRA network of gravitational wave observatories \cite{EventHorizonTelescope:2019dse, EventHorizonTelescope:2021srq, LIGOScientific:2016aoc}.

Below we consider bare black holes, so black holes in vacuum except for the test charges. We are not imagining black holes immersed in a magnetosphere, which would lead to the force-free condition $\bt{E} \cdot \bt{B} =0$, as in \cite{Blandford:1977ds}. We review the Wald argument for spinning black holes in order to show that dynamics upstage screening. (An alternative motto might be ``initial conditions matter.'')  In brief, accounting for gravitational effects turns the Wald charge into a band of stable charges, as $Q$ must be sufficiently above or below the Wald charge for the electrical repulsion to overcome the gravitational attraction. We then dive into the dynamics around boosted black holes to demonstrate charge acquisition as well as chaos.

\subsection{Spinning Black Holes and Screened Electric Fields}
\label{sec:Wald}

In a classic paper from 1974, Wald argued that to screen the electric field generated by the spin $a$ of a black hole in a uniform magnetic field, black holes would acquire a charge $Q_W = 2 a M B_0$, hereafter called the Wald charge \cite{Wald:1974np}. A spinning black hole of charge $Q_W$ will be screened in the sense that charges will not experience a drop in the electromagnetic potential between infinity and the event horizon. Wald's argument was formulated on the spin axis. In a recent article, we showed that the argument could be extended off-axis, although charges will still flux and radiate around the black hole \cite{Levin:2018mzg}. 

The gravito-electrodynamics can lead to charge acquisition that deviates from the value anticipated from screening arguments, $Q_W$. For this purpose it is sufficient to consider particles along the spin axis. We take the spin of the black hole and the magnetic fields to be aligned with the $z$-axis.

In the spacetime described above, the electromagnetic energy per unit mass of a particle with charge per unit mass $\ol{q}$ is
\be \ol{q} A_t = -\chi_Q(g_{tt} + 1). \ee
The final term in the potential is a gauge term we have subtracted to ensure that the electrostatic potential is zero at infinity, as in \cite{Gupta:2021vww}, and we have defined 
\be \chi_Q = \ol{q} \left(\frac{Q}{2M} - aB_0\right) \ee
as in \cite{Gupta:2021vww}. On the $z$-axis,
\be g_{tt} = -\frac{1}{g_{rr}} = -\frac{\D}{\S},\ee
with 
\begin{align}
    \D &= z^2 - 2Mz + a^2 \nonumber \\
    \S &= z^2 + a^2.
\end{align}
Thus the conserved energy per unit mass for the particle is given by 
\be
    e = -\frac{1}{m}\p_\m \h^\m 
    = -(u_t + \ol{q} A_t),
\ee
where $\p^\m = m u^\m + q A^\m$ is the canonical momentum and $\h^\m$ is the timelike Killing vector. The change in the electromagnetic potential for a particle falling from infinity to cross the black hole event horizon at $r_+$ is then
\begin{equation}
    \D\e = \left.
    -\ol{q} A_t\right |_{z=r_+} 
    + \left.\ol{q} A_t \right|_{z=\infty} 
    = \chi_Q
\end{equation}
which vanishes at the Wald charge $Q_W = 2a M B_0$. Consequently, the electric field is screened at the Wald charge \cite{Wald:1974np, DOrazio:2015jcb, King:2021jlb, Komissarov:2021vks}.

However, the dynamics can override the desire to screen the electric field. If we seed particles initially at rest along the $z$-axis stretching from the event horizon out to infinity, the value of the charge that precipitates onto the black hole depends on the initial data which we will now show.

On the positive $z$-axis, $\th = 0$, and the equations of motion simplify greatly with only two coordinates, $(t, z)$. We can use an effective potential formulation since there exist enough constants of motion, $e$ and $u\cdot u$, to reduce the ODEs to first order equations. From $e$ we have 
\begin{equation}
    \dot t = \frac{\S}{\D}(e - \chi_Q) + \chi_Q.
    \label{Eq:tdot}
\end{equation}
Now we leverage $u\cdot u =-1$, using \eqref{Eq:tdot} to eliminate $\dot t$ to write 
\begin{equation}
    \frac{1}{2}\dot z^2 + V_{\rm eff}(z)
    = \frac{e^2-1}{2}
    \label{Eq:effeq}
\end{equation}
with
\begin{equation}
    V_{\rm eff}(z) = \frac{2Mz}{\S}\chi_Q 
    \left(e - \chi_Q \left(\frac{Mz}{\S}\right)\right) - \frac{Mz}{\S}.
\end{equation}
As $z \to \infty$, $V_{\rm eff}\to 0$ and therefore for $\dot z_0 = 0$ we have $e = 1$ in this limit. Notice for this value that $\dot t_0 = 1$. 

\begin{figure}[t]
\centering
\begin{tabular}{>{\centering\arraybackslash} m{5em}|c|c}
& $\left|\chi_Q\right| = 0.5$ & $\left|\chi_Q\right| = 1.5$  \\ \hline
\begin{tabular}{l}
$z_0 = 2.5 M$
\end{tabular} &
\begin{tabular}{l}
\includegraphics[width=.3\linewidth]{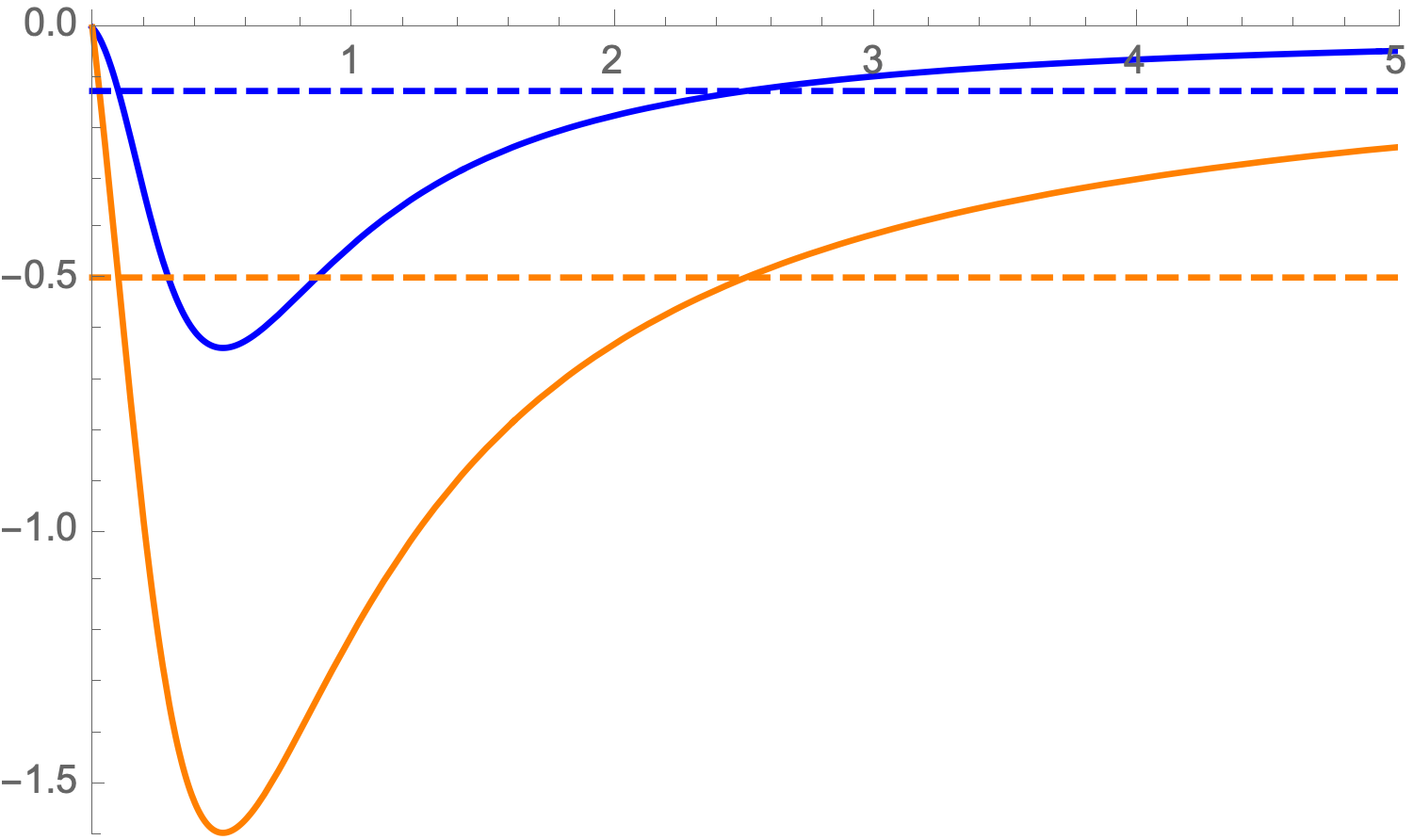}
\end{tabular} &
\begin{tabular}{l}
\includegraphics[width=.3\linewidth]{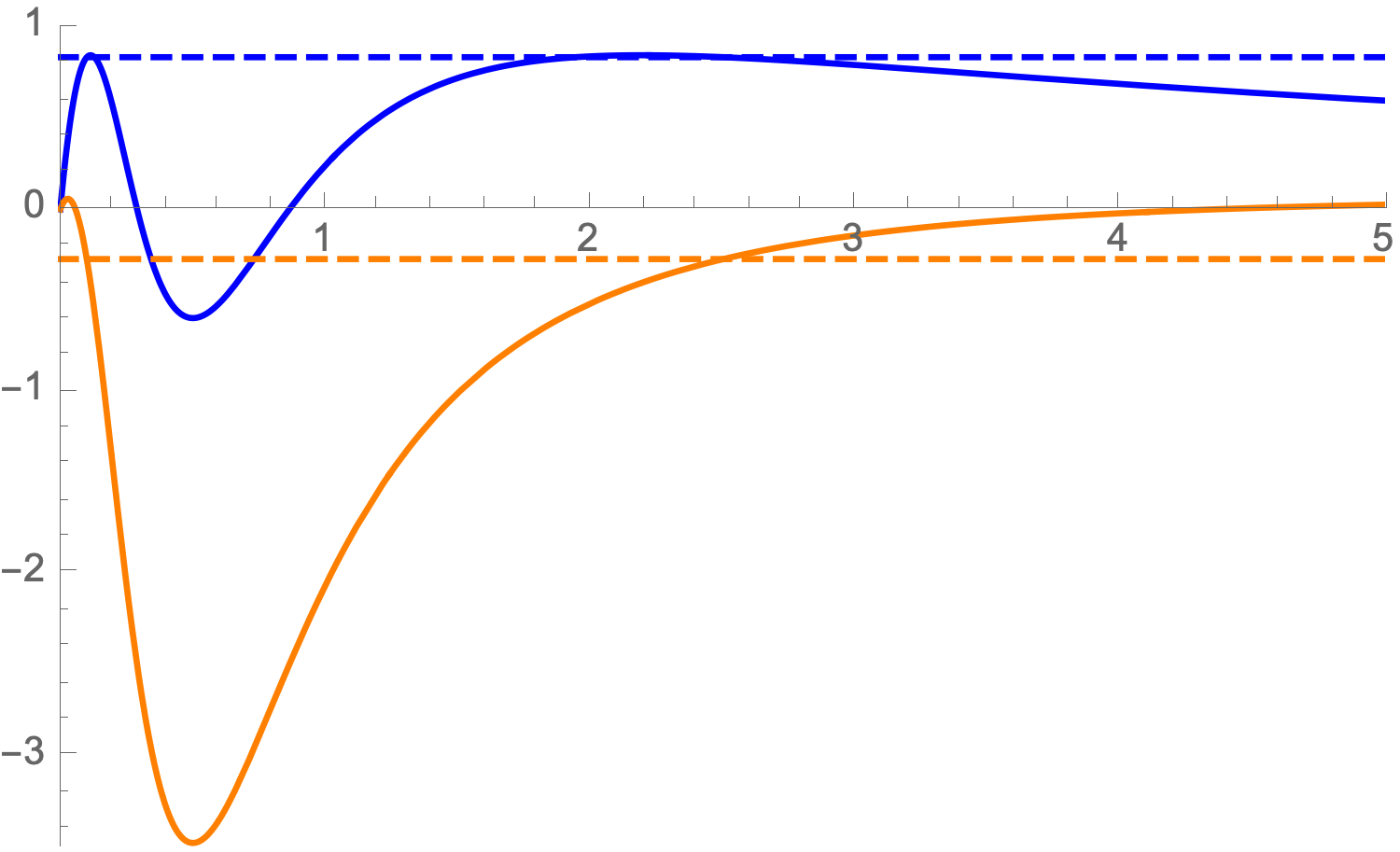}
\end{tabular}
\\
\hline
\begin{tabular}{l}
$z_0 = 5 M$
\end{tabular} &
\begin{tabular}{l}
\includegraphics[width=.3\linewidth]{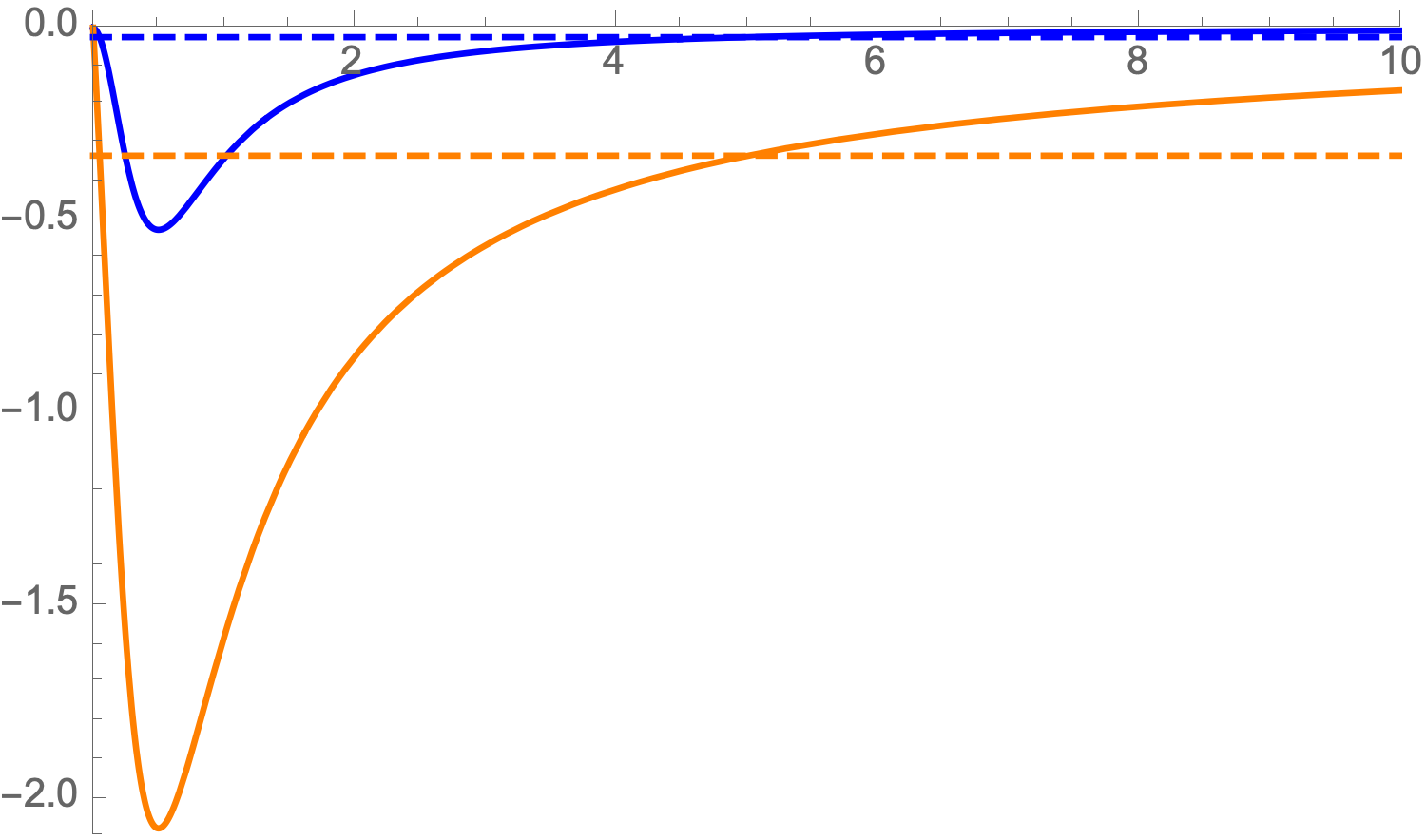}
\end{tabular} &
\begin{tabular}{l}
\includegraphics[width=.3\linewidth]{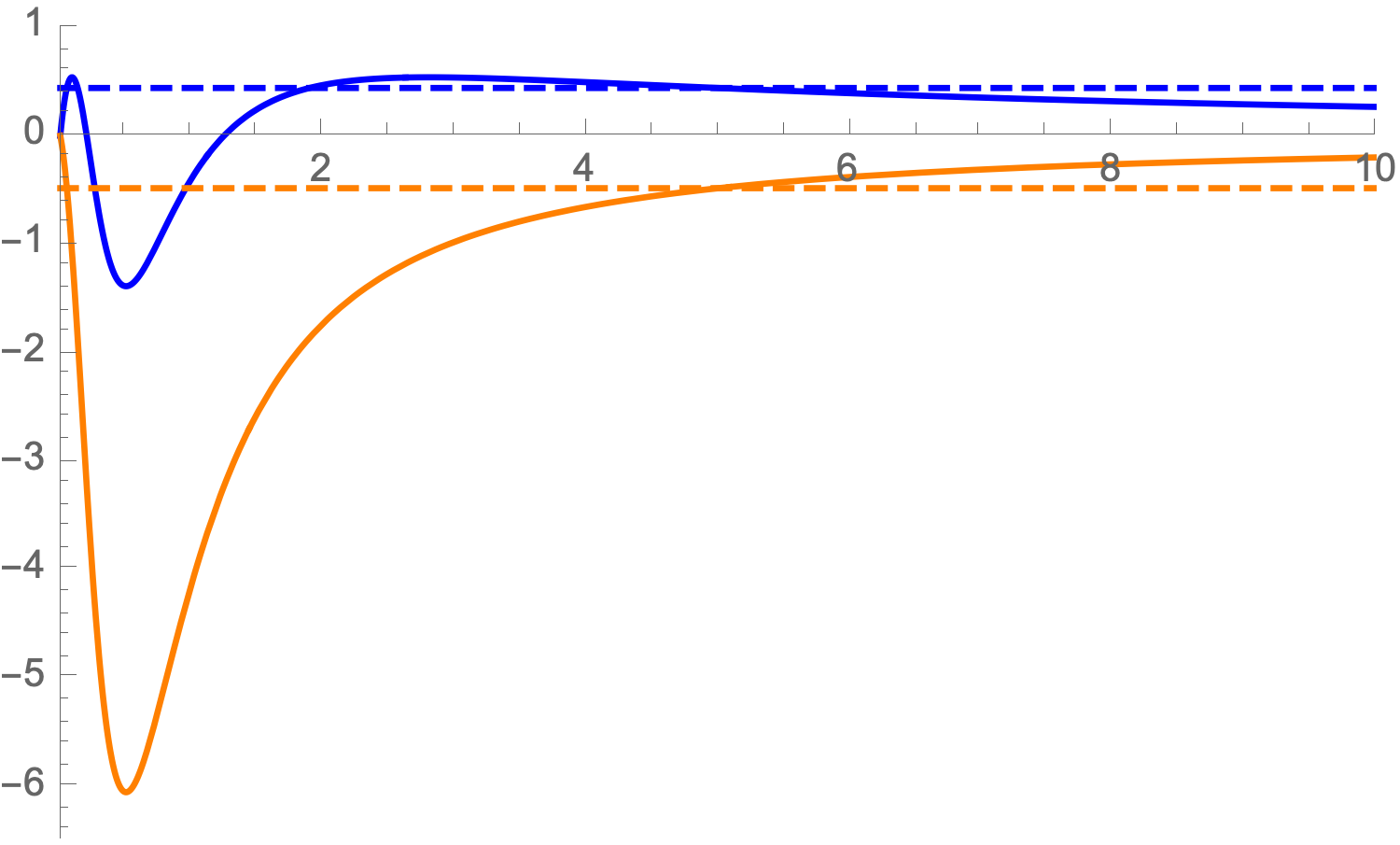}
 \end{tabular}
\\
\end{tabular}
\caption{Sample effective potentials with $a = 0.5M$ with varying $z_0$ and $Q$. Solid lines correspond to effective potential and dashed lines to the energy of a charged particle starting at rest at the listed $z_0$. The blue plots have $\chi_Q > 0$ and the orange have $\chi_Q < 0$. For fixed $z_0$ and $\chi_Q > 0$, the effective potential increases for large $z$ for small $\chi_Q$ and decreasing for larger $\chi_Q$. For $z_0 = 2.5M$ the transition between the two behaviors occurs at $\chi_Q \approx 1.4$, and for $z_0 = 5M$ at $\chi_Q \approx 0.64$.}
    \label{fig:random_veffs}
\end{figure}

Letting $\S_0$ and $\D_0$ denote the values of $\S$ and $\D$ at $z = z_0$, we set $\dot{z}_0 = 0$ so that the  particles are initially at rest, finding $\dot t_0 = \sqrt{\frac{\S_0}{\D_0}}$, which constrains
\begin{equation}
    e = \frac{\D_0}{\S_0} \left(\sqrt{\frac{\S_0}{\D_0}} - \chi_Q\right) + \chi_Q.
\end{equation}
As required, $e$ approaches 1 as $z_0 \to \infty$. We can use this value for $e$ in $V_{\rm eff}(z)$ as shown in Fig.\ \ref{fig:random_veffs}.

As the potential diagrams show, the dynamics can favor one sign of charge over another even away from $\chi_Q = 0$. Since the slope of the effective potential at $z_0$ determines whether the particle moves towards or away from the black hole originally, we solve for $\chi_Q^*$, the value at which $V_\text{eff}'(z_0) = 0$:
\be
\chi_Q^* \equiv \frac{1}{2}\sqrt{\frac{\S_0}{\D_0}}.
\ee
We find $V_\text{eff}'(z_0) > 0$ for $\ch_Q < \ch_Q^*$ and $V_\text{eff}'(z_0) < 0$ for $\ch_Q > \ch_Q^*$. 

It turns out that the sign of $V_\text{eff}'(z_0)$ is sufficient for determining the subsequent behavior of the particles, which can be seen by finding the zeros of $V_\text{eff}'(z)$. Ignoring the roots at $z = \pm a$, we find that for $\ch_Q > \ch_Q^*$, one of the other two roots is less than $r_+$ and the other is less than $z_0$, so $V_\text{eff}'(z) < 0$ for $z \geq z_0$, and thus the particles will go out to infinity. For $0 < \chi_Q < \chi_Q^*$, either the two roots are complex, or one is less than $r_+$ and the other is now greater than $z_0$. Thus $V_\text{eff}'(z) > 0$ for $r_+ \leq z \leq z_0$, so the particles will fall into the black hole.

We see that $\chi_Q^*$ goes to $1/2$ as $z_0 \to \infty$. In other words, above the Wald charge, positive charges are still favored, and below the Wald charge, negative charges are still favored in a band defined by
\begin{equation}
    Q_W - \frac{M}{\left|\ol{q}\right|} < Q < Q_W + \frac{M}{\left|\ol{q}\right|}.
\end{equation}

\begin{figure}[tb]
    \centering
        \begin{tabular}{cc}
        \includegraphics[width=.45\linewidth]{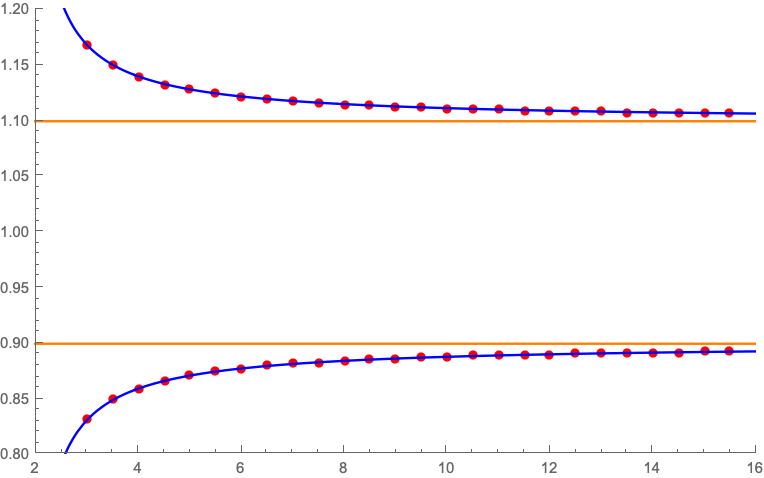} &  \includegraphics[width=.3\linewidth]{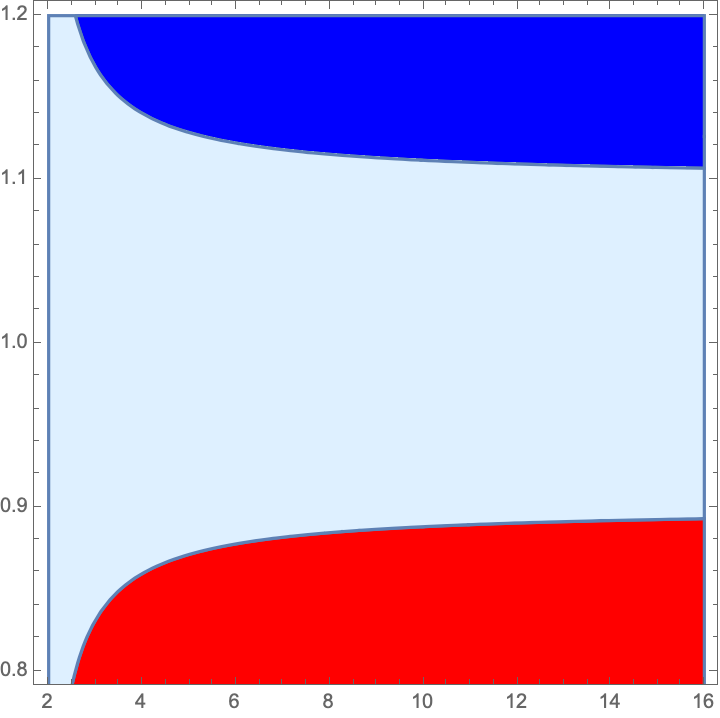} \\
    \end{tabular}
    \caption{Left: $Q/Q_W$ vs. $z_0/M$. The blue curves are $Q_W \pm \frac{M}{|\ol{q}|}\sqrt{\frac{\S_0}{\D_0}}$ in units of $Q_W$. The straight orange lines are the asymptotic values $Q_W \pm \frac{M}{|\ol{q}|}$. The red dots are the results of the simulations. The initial charge of the black hole was taken to be above the blue curve for the simulations yielding dots above the band, and vice versa for the dots below the band. The parameters are $\ol{q} B_0 = 10 \, M^{-1}$ and $a = 0.5 M$, so $Q_W = 0.1 M$. Right: The same plot, but this time with each each region color-coded based on the behavior of test particles. Blue corresponds to only negative particles falling, red to only positive falling, and light blue to both particles falling in.}
    \label{fig:WaldPlot}
\end{figure}

To examine this behavior, we simulated the trajectories of a series of particles. We seeded positive and negative particles at $z_0$ and let them evolve. If both particles escaped or both fell into the black hole, the simulation ended. If one fell in and one escaped, the charge of the black hole was increased or decreased and the point was seeded again, repeating until the charge of the black hole stabilized. For the first set of simulations, we took the initial charge to obey $Q_0 > Q_W + \frac{M}{|\ol{q}|}\sqrt{\frac{\S_0}{\D_0}}$, and took $Q_0 < Q_W - \frac{M}{|\ol{q}|}\sqrt{\frac{\S_0}{\D_0}}$ for the second set. Fig.\ \ref{fig:WaldPlot} shows the final charge in units of $Q_W$ as a function of $z_0$ in units of $M$.

Fig.\ \ref{fig:WaldPlot} can best be understood by thinking of the blue curves as the boundaries of regions in the $z_0$-$Q_0$ plane that determine the behavior of the charged particles. Above the upper blue curve, positive charges seeded at rest at $z_0$ move away from the black hole, and below the upper below curve they fall in. Below the lower blue curve, negative particles move away from the black hole, and above the lower blue curve they fall in. Thus in the region between the curves, particles of both charge fall in, so the charge of the black hole is stable. Essentially, accounting for gravitational effects turns the Wald charge into a band of stable charges, as $Q$ must be sufficiently above or below $Q_W$ for the electrical repulsion to overcome the gravitational attraction. 
This idea of overlapping regions in the $z_0$-$Q_0$ plane that define the behavior of the particles will be useful in Section 3.

The nearer you are to the black hole, the larger the deviation as shown in Fig.\ \ref{fig:WaldPlot}. This result is in no way generalizable. Some parameter values will lead to oscillatory orbits and significant deviations from the figure. We stress that we are pressing the point that the specific initial seeding of charges and the gravito-electrodynamics can favor accretion of charges of a particular sign, even if this does not lead to the full screening effect that Wald originally described. Even more, we have quantified the stable band around the charged black hole analytically.

We take that lesson with us as we leave spinning black holes behind and turn our attention to boosted black holes.

\section{Maxwell Equations around a Boosted Black Hole}

We take the black hole to be immersed in an external magnetic field that is asymptotically uniform in the positive $z$ direction. We also take the black hole to be moving with constant speed $\b$ in the negative $y$ direction. In the frame of the black hole, the magnitude of the asymptotic field is $B_0$ and there is an asymptotic electric field in the negative $x$ direction, $\bt{E}_0 = \bs{\b} \times \bt{B}_0$. In Appendix A, we find the value of the electromagnetic fields everywhere. 

The Schwarzschild metric is given by
\begin{equation}
ds^2 = -N^2 c^2\, dt^2 + N^{-2} \, dr^2 + r^2 \, d\O^2,
\end{equation}
with $N^2 = 1 - \dfrac{2M}{r}$. The Maxwell equations in vacuum are
\begin{equation}
\del_\n\left(\sqrt{-g}F^{\m\n}\right) = 0, \quad
\del_\n\left(\sqrt{-g} (\star F)^{\m\n}\right) = 0,
\label{Maxwell}
\end{equation}
where $F_{\m\n} = \del_\m A_\n - \del_\n A_\m$ and $(\star F)^{\m\n} = \frac{1}{2} \sqrt{-g}\e^{\m\n\r\s} F_{\r\s}$. We neglect the contribution of the electromagnetic fields to the geometry of the spacetime. Allowing for charge on the black hole, we solve the Maxwell equations, finding
\begin{align}
A_t &= -\frac{Q}{r} 
- E_0\left(1 - \frac{2M}{r}\right)
r \sin\th\cos\f\\
A_\f &= \frac{1}{2} B_0 r^2 \sin^2\th.
\label{eq:magpotphi}
\end{align}
as did the authors of \cite{Morozova:2013ina} previously.

\begin{figure}[h]
\begin{tabular}{ccc}
$Q = 0$ & $Q = 0.5 M$ & $Q = M$ \\
\begin{tabular}{l}
\includegraphics[width=.3\linewidth]{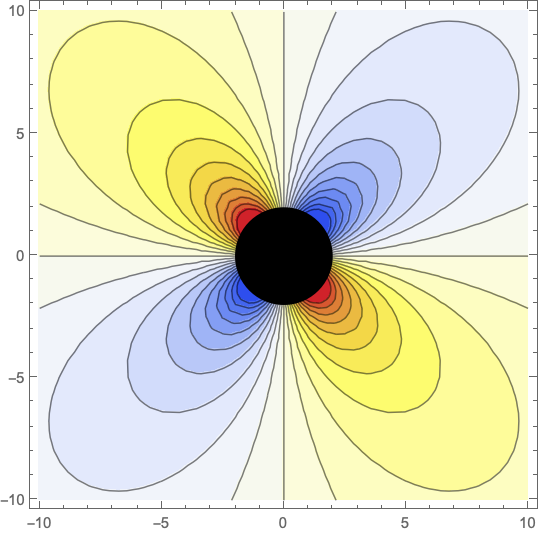}
\end{tabular} &
\begin{tabular}{l}
\includegraphics[width=.3\linewidth]{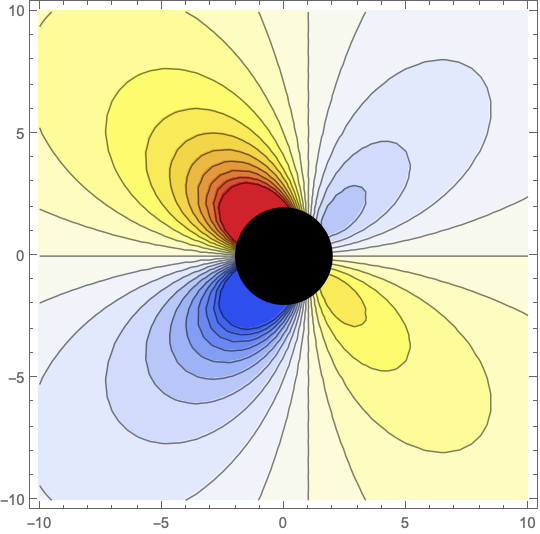}
 \end{tabular} &
\begin{tabular}{l}
\includegraphics[width=.3\linewidth]{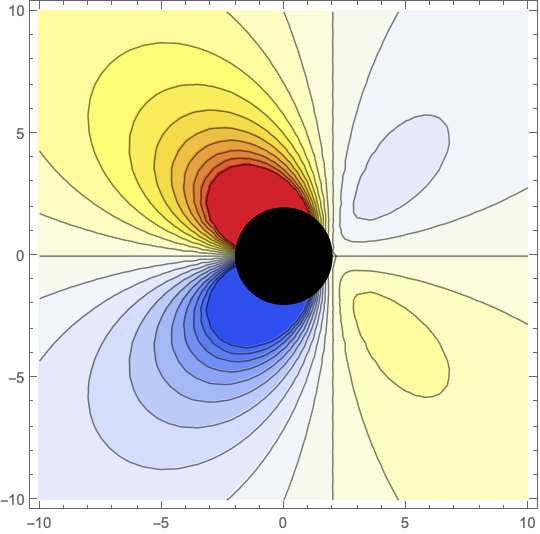}
\end{tabular}\\
\multicolumn{3}{c}{\includegraphics[width=.7\linewidth]{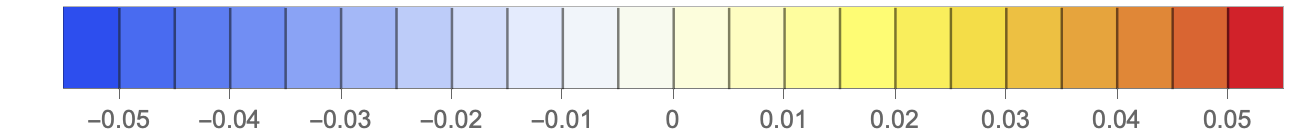}}
\end{tabular}
\caption{Plots of $\bt{E} \cdot \bt{B}$ in the $y$-$z$ plane in units of $M^{-2}$ for $\ol{q} B_0 = 0.5 M^{-1}$, $\b = 0.5$, and varying $Q$.}
\label{fig:EdotB}
\end{figure}

For reference, we also report the invariant,
\begin{align}
\bt{E} \cdot \bt{B} 
&= \frac{1}{4}F^{\m\n} (\star F)_{\m\n}\nonumber\\
&= \frac{Q B_0}{r^2}\cos\th 
- E_0 B_0 \frac{2M}{r}\cos\th \sin\th \cos\f.
\label{eq:edotb}
\end{align}
Derivations of \eqref{eq:edotb} and \eqref{eq:magpotphi} can be found in Appendix A.
Contour plots of $\bt{E} \cdot \bt{B}$ in the $y$-$z$ plane for various values of $Q$ are shown in Fig.\ \ref{fig:EdotB}. We note that this vanishes in the equatorial plane. Even with a force-free magnetosphere, the best charges can do in attempts to screen the electric field is to achieve $\bt{E} \cdot \bt{B}=0$. In our vacuum scenario, there is no value of $Q$ that leads to the force-free condition $\bt{E} \cdot \bt{B} = 0$ everywhere. In the recent Ref.\ \cite{Chen:2021sya}, the authors considered the electromagnetic field structure to motivate charged boosted black holes from asymmetric distributions.

\subsection{Screening of the electric field}

Although we expect charge acquisition to deviate substantially from a prediction based on screening of the electric field, it is worth considering the effect of $Q$ on the electric potential, in analogy with section \ref{sec:Wald} for a spinning black hole. 

We consider a test particle of charge per unit mass $\ol{q}$ moved from $r = r_0$ to the horizon. As before, the conserved energy per mass of the particle is 
\begin{equation}
e = -(u_\m + \ol{q} A_\m) \h^\m
\end{equation}
Following Wald, we see that the change in electrostatic energy per unit mass is
\begin{equation}
\D\e = -\ol{q} A_\m \h^\m\big\vert_{r = 2M} 
+ \ol{q} A_\m \h^\m\big\vert_{r = r_0}
\end{equation}
or
\begin{equation}
\D\e = \frac{\ol{q} Q}{2M} 
- \frac{\ol{q} Q}{r_0}
- \ol{q} E_0 N(r_0)^2 r_0 \sin\th\cos\f.
\end{equation}
We can rewrite this as
\begin{equation}
\D\e = \frac{\ol{q}}{2M} 
\left (Q - Q_T(\theta,\f)\right )\left ( 1-\frac{2M}{r_0}\right )
\end{equation}
with the threshold charge
\begin{equation}
   Q_T(\th,\f)=2M E_0 r_0  \sin\th\cos\f.
\end{equation}
Notice that $Q_T$ averages to zero. Clearly, and not surprisingly, there is no value of $Q$ that screens the electric field in the frame of the black hole.

The electric field does disappear in the frame of test charges that are at rest with respect to $\g^{-1} B_0$ (for an uncharged black hole). However, as the black hole whizzes by, these charges will automatically begin to move and thereby will begin to feel an electric field. The gravito-electrodynamics then determine the acquisition of charge on the black hole and not the aspiration to screen the electric field.

\section{The dynamics of charging up a boosted black hole} 

For concreteness, we set particles at rest with respect to each other until the black hole flies by. In the frame of the black hole, the charges have initial velocity $\bt{v}_0 = \b\hbt{y}$. More precisely, this is the velocity measured by a local observer at the initial location of the particle. In Appendix B, we define a proper orthonormal basis to explicitly set the initial data with relativistic effects included.

\begin{figure}[tb]
    \centering
    \includegraphics[width=0.9\linewidth]{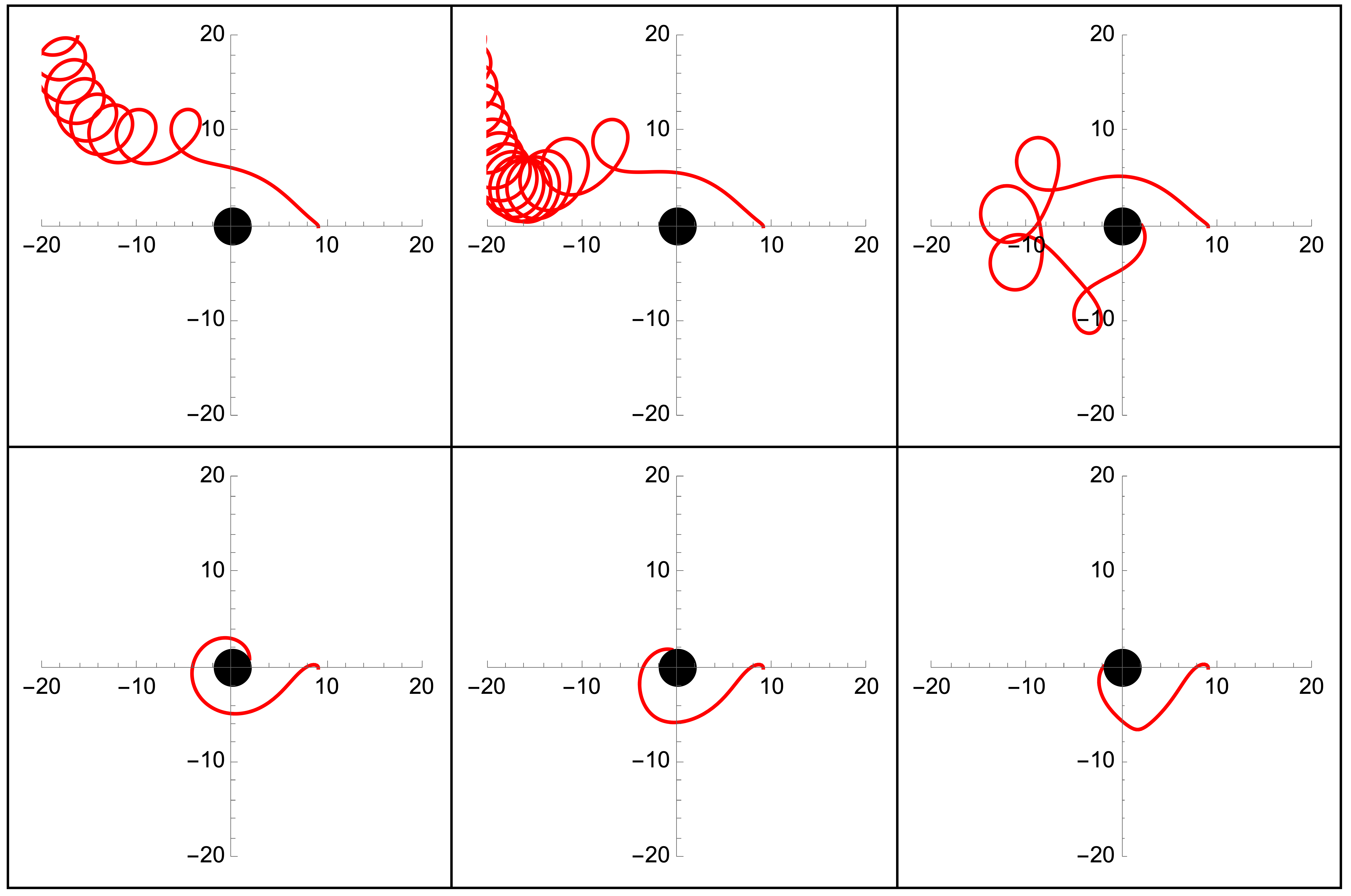}
    \caption{Plot of charged particle trajectories for $\ol{q} B_0 = 0.1 M^{-1}$, $\b = 0.05$, and $r_0 = 9M$. The top plots have positive charges and the bottom plots have negative charges. Each column is at a fixed $Q$, with $Q$ decreasing from left to right by $0.1 M$ until the final column where both the positive and negative test charge fall in, stabilizing the black hole charge at $Q = -0.3M$. The $x$ and $y$ coordinates are in units of $M$.}
    \label{fig:regionII_trajectories}
\end{figure}

Intuitively, we expect that test particles extremely close to the event horizon will indiscriminately fall in, unless charges are absurdly high.
Similarly, far enough away from the black hole, the dynamics of test charges will be dominated by electromagnetic forces.
In the intermediate region within tens of Schwarzschild radii, the gravito-electrodynamics are quite complicated. In fact, the dynamics are most assuredly chaotic as we have lost two constants of motion, both the angular momentum and the Carter constant.

The magnetic field selects a preferred equatorial plane around the black hole. Only trajectories in the equatorial plane remain two-dimensional. To explore the dynamics of charged particles and the influence on charge acquisition, we will pay particular attention to these equatorial orbits. Furthermore, the electric field picks out the $x$-axis as special. We begin by exploring orbits that are seeded on the $x$-axis with the initial data set as described above. We then consider orbits seeded anywhere in the equatorial plane.
 
 \begin{figure}[tb]
    \centering
    \includegraphics[width=0.9\linewidth]{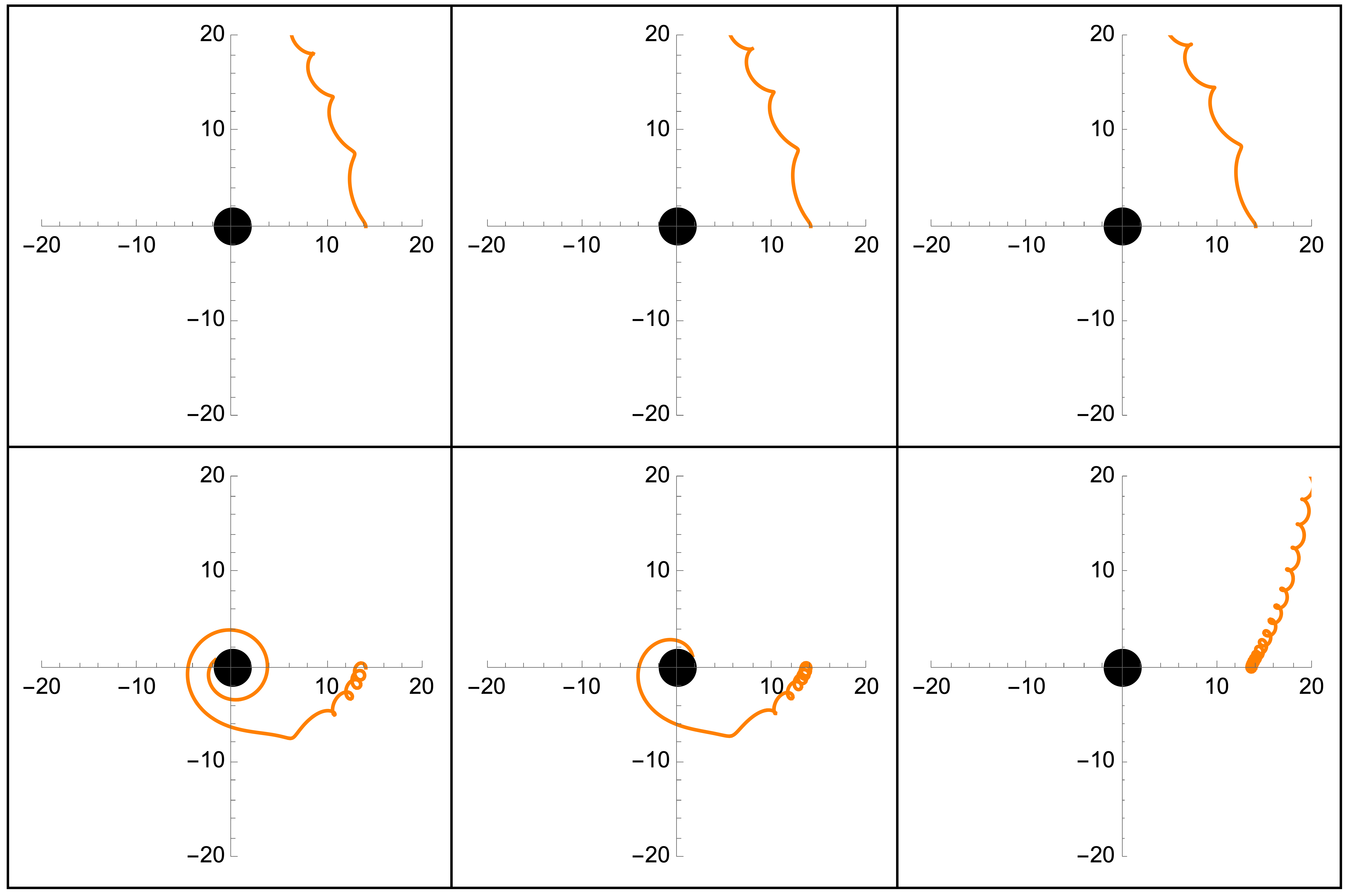}
    \caption{
    Plot of charged particle trajectories for $\ol{q} B_0 = 0.1 M^{-1}$, $\b = 0.05$, and $r_0 = 14M$. The top plots have positive charges and the bottom plots have negative charges. Each column is at a fixed $Q$, with $Q$ decreasing from left to right by $0.1 M$ until the final column where both the positive and negative test charge escape, stabilizing the black hole charge at $Q = -0.2M$. $x$ and $y$ are in units of $M$.}
    \label{fig:regionIII_trajectories}
\end{figure}

\subsection{Orbits seeded on the $x$-axis}

\begin{table}[!ht]
    \centering
\begin{tabular}{ |c|c| } 
 \hline
$r_0 < r_I$ & Both charges fall in. \\ \hline
$r_I < r_0 < r_{II}$ & Negative charges fall in until a positive charge falls in. \\ \hline
$r_{II} < r_0 < r_{III}$  & Negative charges fall in until they are repelled. 
Positive charges always escape. \\ \hline
$r_{III} < r_0$ & Both charges escape. \\ \hline
\end{tabular}
    \caption{Summary of behavior of charges in each interval.}
    \label{tab:Bfieldsummary}
\end{table}

We plant charges on the positive $x$-axis at varying $x = r_0$ and observe four different kinds of behavior with ascending radius as summarized in the chart above. The exact values of the radii at which we transition between the four behaviors depends on the parameters $\b$ and $\ol{q}B_0$. 

For radii $r_0 < r_I$, both charges fall in and the black hole remains uncharged. For radii $r_I< r_0 < r_{II}$, the black hole acquires negative charge. Although positive charges are pulled to the left by the electric field, they overshoot the black hole and so do not contribute to the charge. Consequently, positive charges tend to escape while negative charges fall in. As each particle falls in we update the charge of the black hole, which becomes increasingly negative until a positive charge falls in. This is shown in Fig.\ \ref{fig:regionII_trajectories}.

As we move to larger radii, $r_{II}< r_0 < r_{III}$, the behavior changes. Initially, positive charges still escape and negative charges fall in. But now the charge stabilizes when a negative charge is repelled, not when a positive charge is pulled in. In addition, the negative charges undergo many tight spirals at the beginning of their trajectories. This behavior is shown in Fig~\ref{fig:regionIII_trajectories}. Lastly, at radii $r_0 > r_{III}$, both charges escape and the black hole remains uncharged. 

\subsubsection{Basin Boundaries in the $x_0$-$Q_0$ plane}

\begin{figure}
    \centering
        \includegraphics[width=\linewidth]{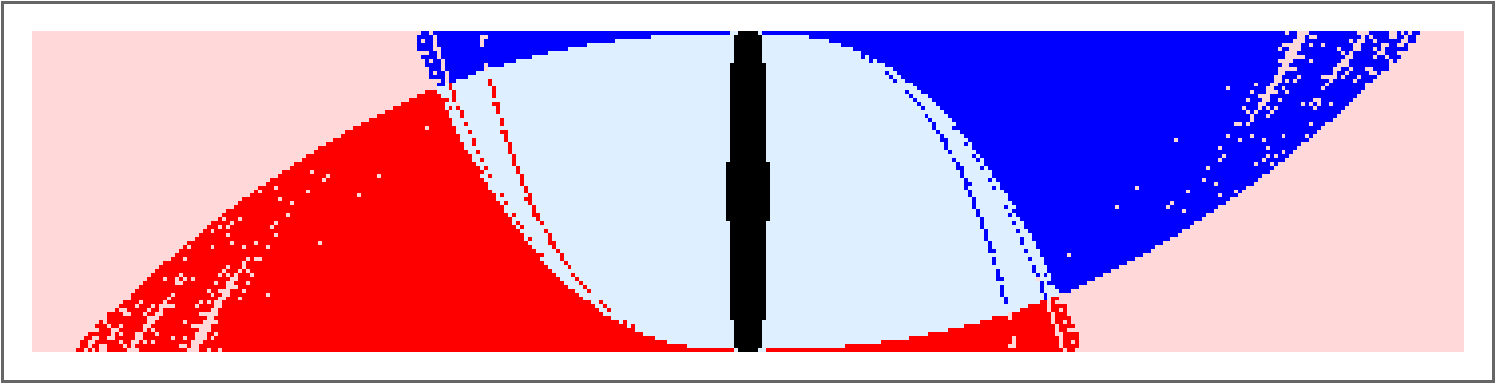} 
    \caption{Plot of the $x_0$-$Q_0$ plane of initial data. $\ol{q}B_0 = 0.01M^{-1}$ and $\b = 0.05$ corresponding to the upper right plot in the grid in Fig.\ \ref{fig:trianglefullgrid}. The $x$ axis has $x_0/M$ from $-72$ to $72$ with $\D x_0/M = 0.4$. The $y$ axis has $Q_0/M$ from $-1$ to $1$ with $\D Q_0/M = 0.025$.}
    \label{fig:Qxplot}
\end{figure}

To better understand the behaviors in these intervals, we perform the same procedure that led to Fig.\ \ref{fig:WaldPlot}, this time in the $x_0$-$Q_0$ plane, leading to figure Fig~\ref{fig:Qxplot}. We simulated the trajectories for a positive and a negative particle for cells throughout the plane, color-coding them as follows:
\begin{itemize}
    \item Black for cells within the black hole. For higher $|Q_0|$, the event horizon shrinks which leads to the tapering at the ends.
    \item Light red if both particles escape, which we see at large $|x_0|$.
    \item Light blue if both particles fall in, which we see at low $|x_0|$.
    \item Blue if only the negative particle falls in.
    \item Red if only the positive particle falls in.
\end{itemize}

As opposed to the plots in Figs. \ref{fig:regionII_trajectories} and \ref{fig:regionIII_trajectories}, here the black hole charge is not updated as the particles fall in. We instead simply color the cell based on the end result for the test particles in the presence of a black hole with specific $Q_0$.

It is helpful to emphasize the connection between Figs.\ \ref{fig:WaldPlot} and \ref{fig:Qxplot}. Again, each charged particle has two regions. Taking one region to be the union of the blue and light blue cells, we see that in this region, negative particles fall in, and outside this region, they escape. Taking the other region to be the union of the red and light blue cells, we see that within this union, positive particles fall in, and outside, they escape. Just as in Fig.\ \ref{fig:WaldPlot}, in the light blue cells (the overlap of these two regions), $Q_0$ is unchanged, because both particles fall in. Unlike in the Wald case, there are points (the light red cells) where both particles escape. A major difference from the Wald case is that the boundaries of these regions are fractals instead of simple curves, as can be seen by the dusty mixing of colors. This is to be expected, as the dynamics are chaotic, unlike the one-dimensional dynamics of the Wald case. 
Another major difference is that the boundaries of these two regions intersect, which explains the behavior seen\footnote{The first column in each figure corresponds to a point on the $x_0$ axis in Fig.\ \ref{fig:Qxplot} where the positive particle escapes and the negative particle falls in, i.e., a blue cell. As the black hole charge is decremented from zero, eventually either the positive particle falls in (Fig.\ \ref{fig:regionII_trajectories}) or the negative particle escapes (Fig.\ \ref{fig:regionIII_trajectories}). The former corresponds to moving downward into the light blue region, and the latter corresponds to moving downward into the light red region.}
in Figs.\ \ref{fig:regionII_trajectories} and \ref{fig:regionIII_trajectories} and summarized in Table\ \ref{tab:Bfieldsummary}.

\subsubsection{Triangles}

\begin{figure}[tb]
\begin{tabular}{>{\centering\arraybackslash} m{5em}|c|c|c}
& $|\ol{q}| B_0 = 0.01 M^{-1}$ & $|\ol{q}| B_0 = 0.05 M^{-1}$ & $|\ol{q}| B_0 = 0.1 M^{-1}$ \\ \hline
\begin{tabular}{l}
$\b = 0.05$
\end{tabular} &
\begin{tabular}{l}
\includegraphics[width=.25\linewidth]{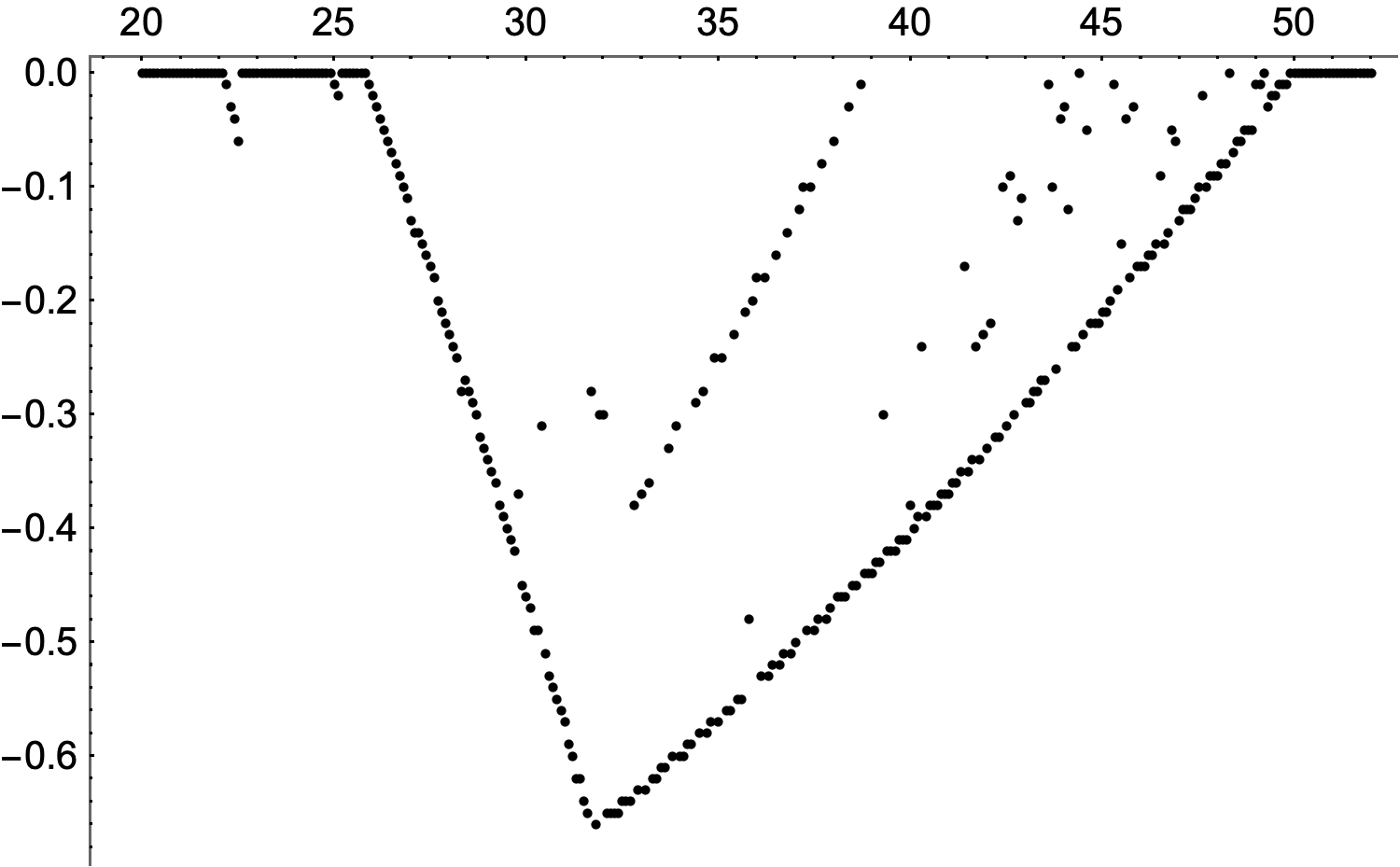}
\end{tabular} &
\begin{tabular}{l}
\includegraphics[width=.25\linewidth]{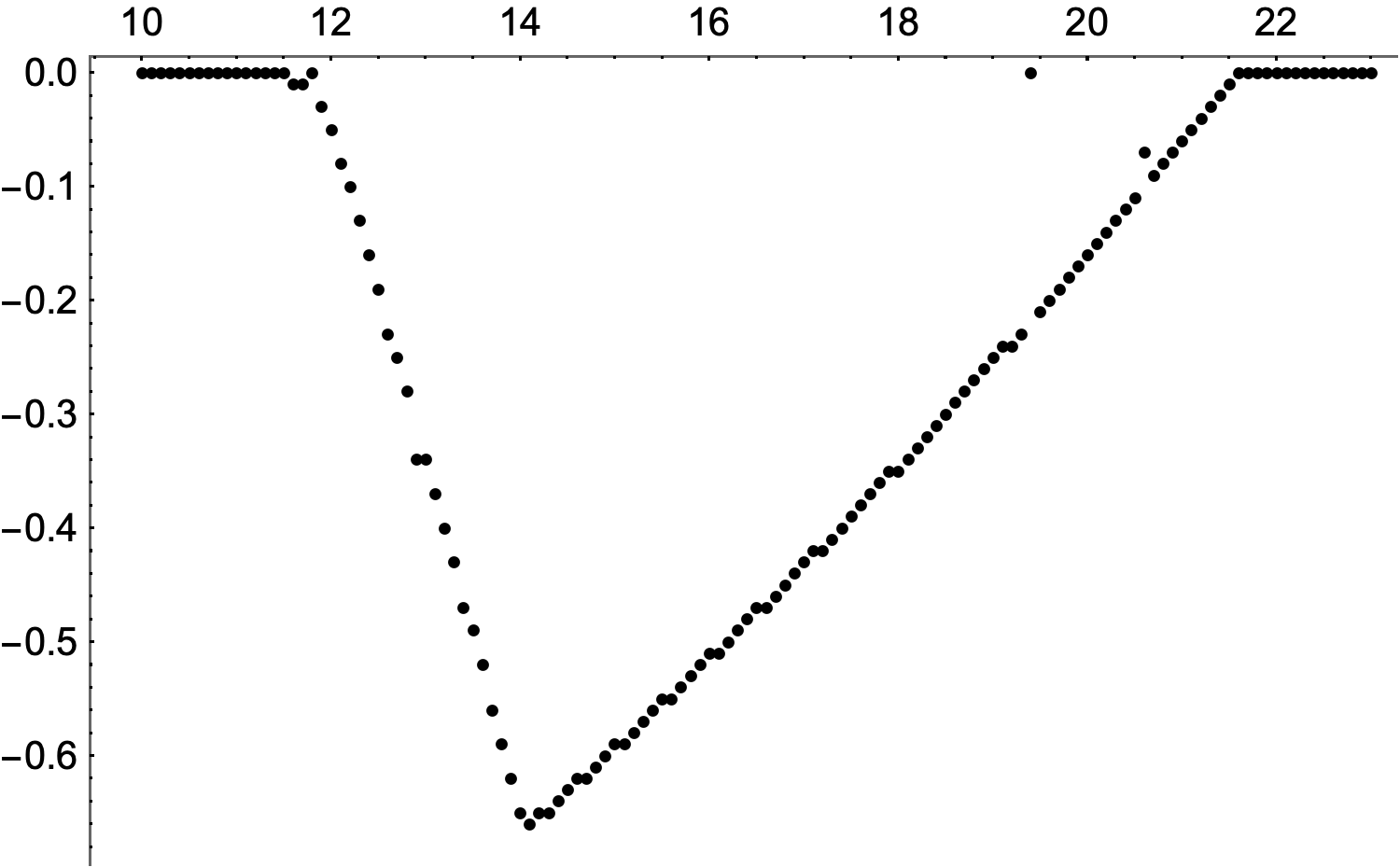}
 \end{tabular} &
\begin{tabular}{l}
\includegraphics[width=.25\linewidth]{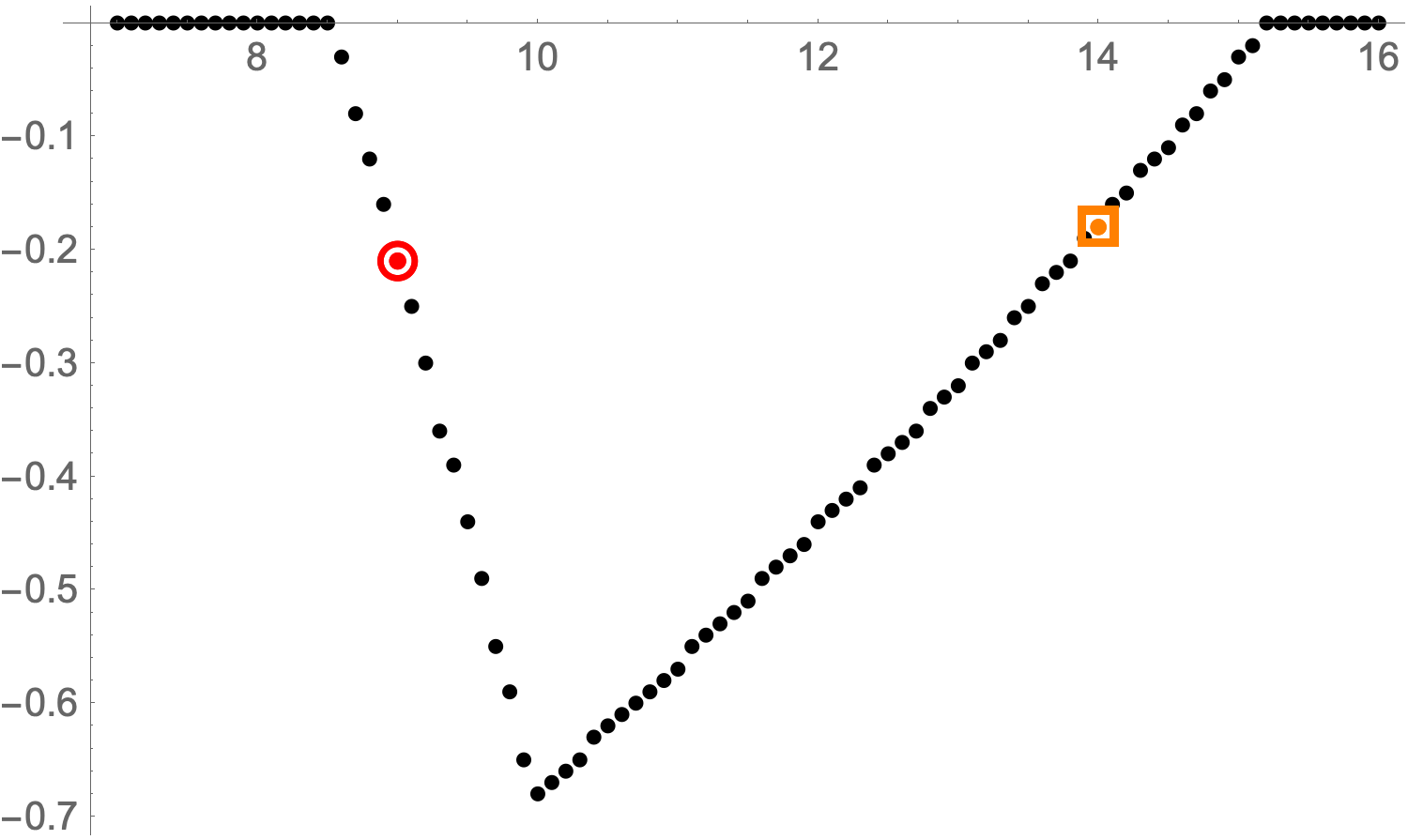}
\end{tabular}
\\
\hline
\begin{tabular}{l}
$\b = 0.15$
\end{tabular} &
\begin{tabular}{l}
\includegraphics[width=.25\linewidth]{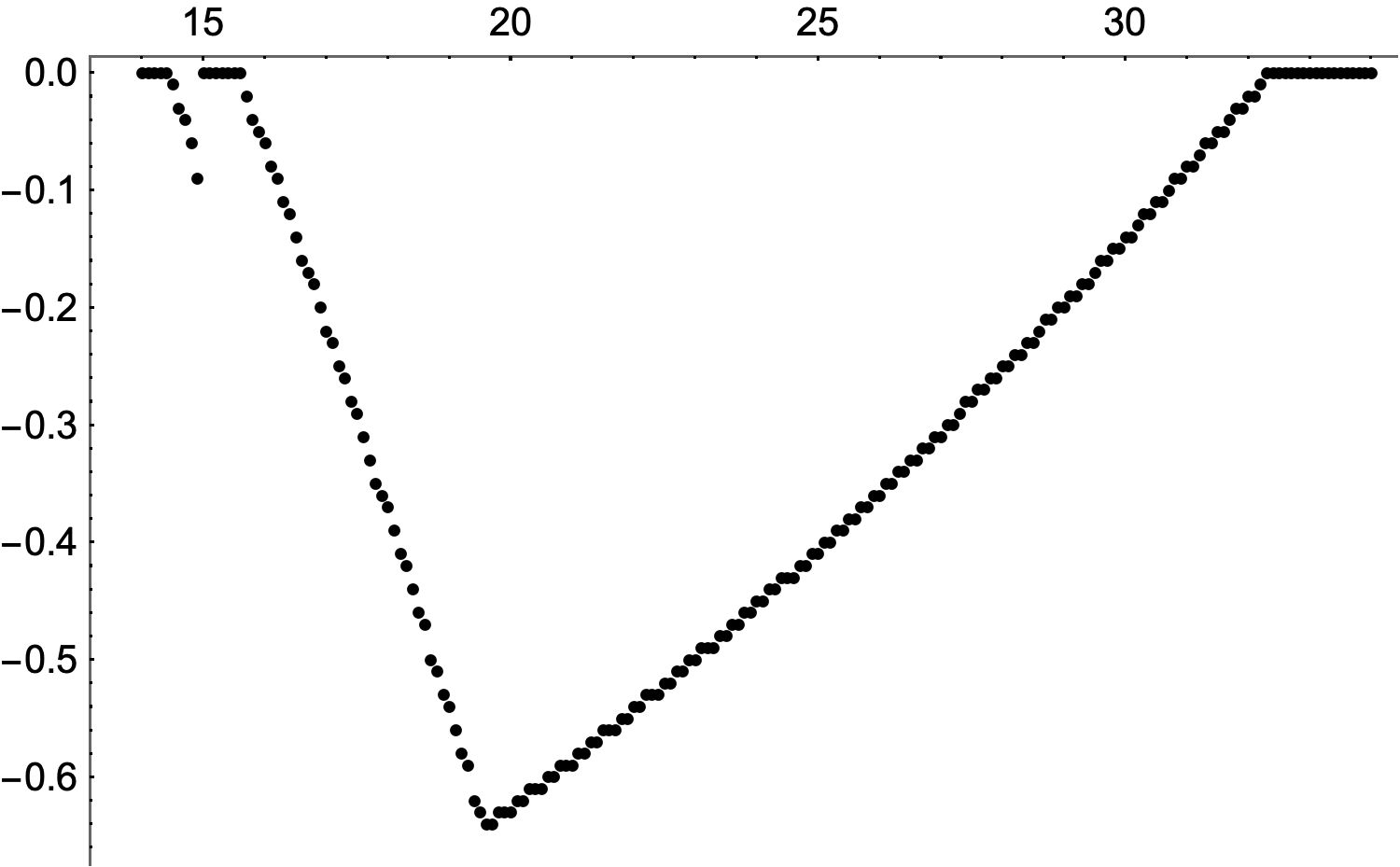}
\end{tabular} &
\begin{tabular}{l}
\includegraphics[width=.25\linewidth]{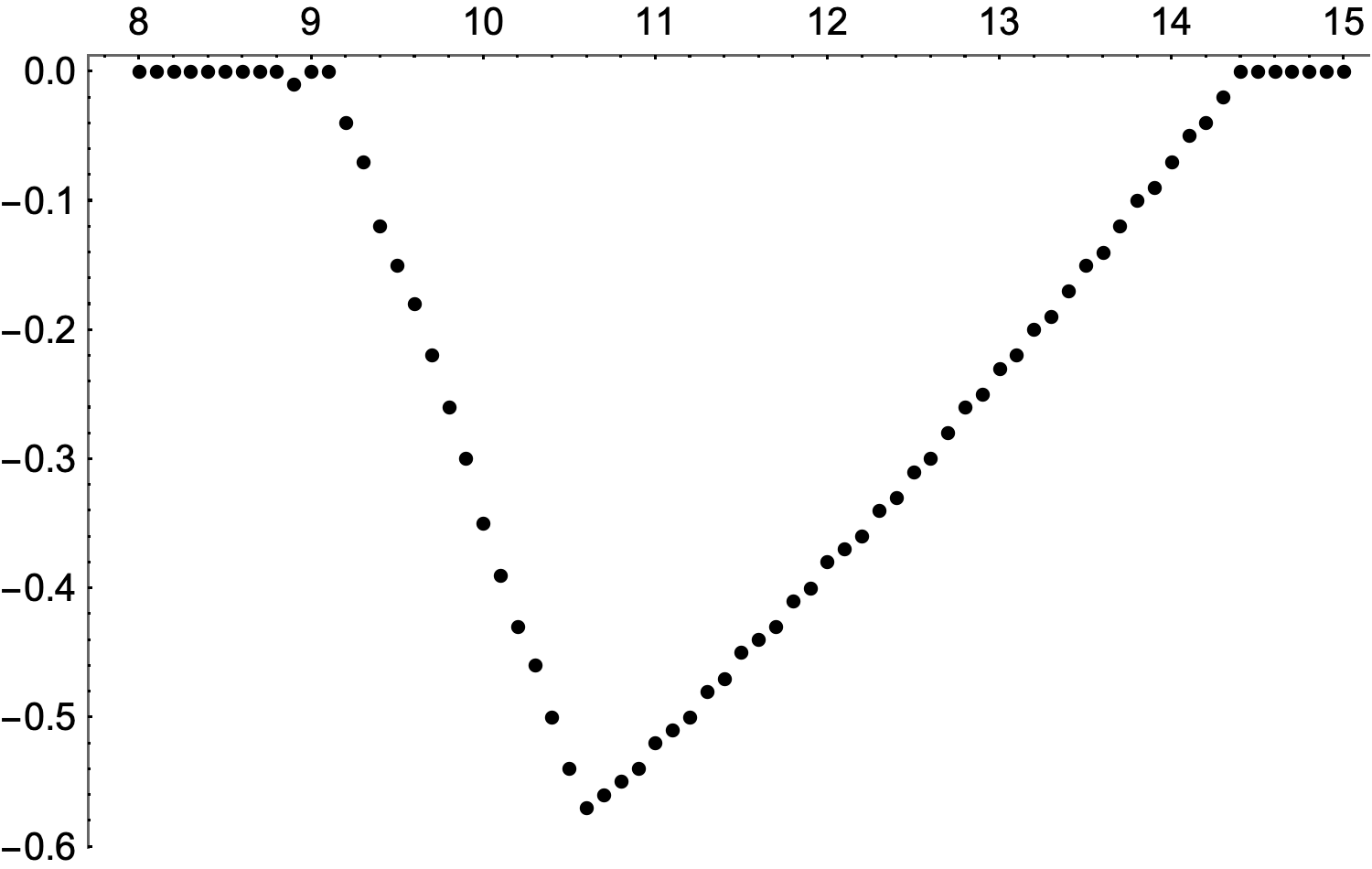}
 \end{tabular} &
\begin{tabular}{l}
\includegraphics[width=.25\linewidth]{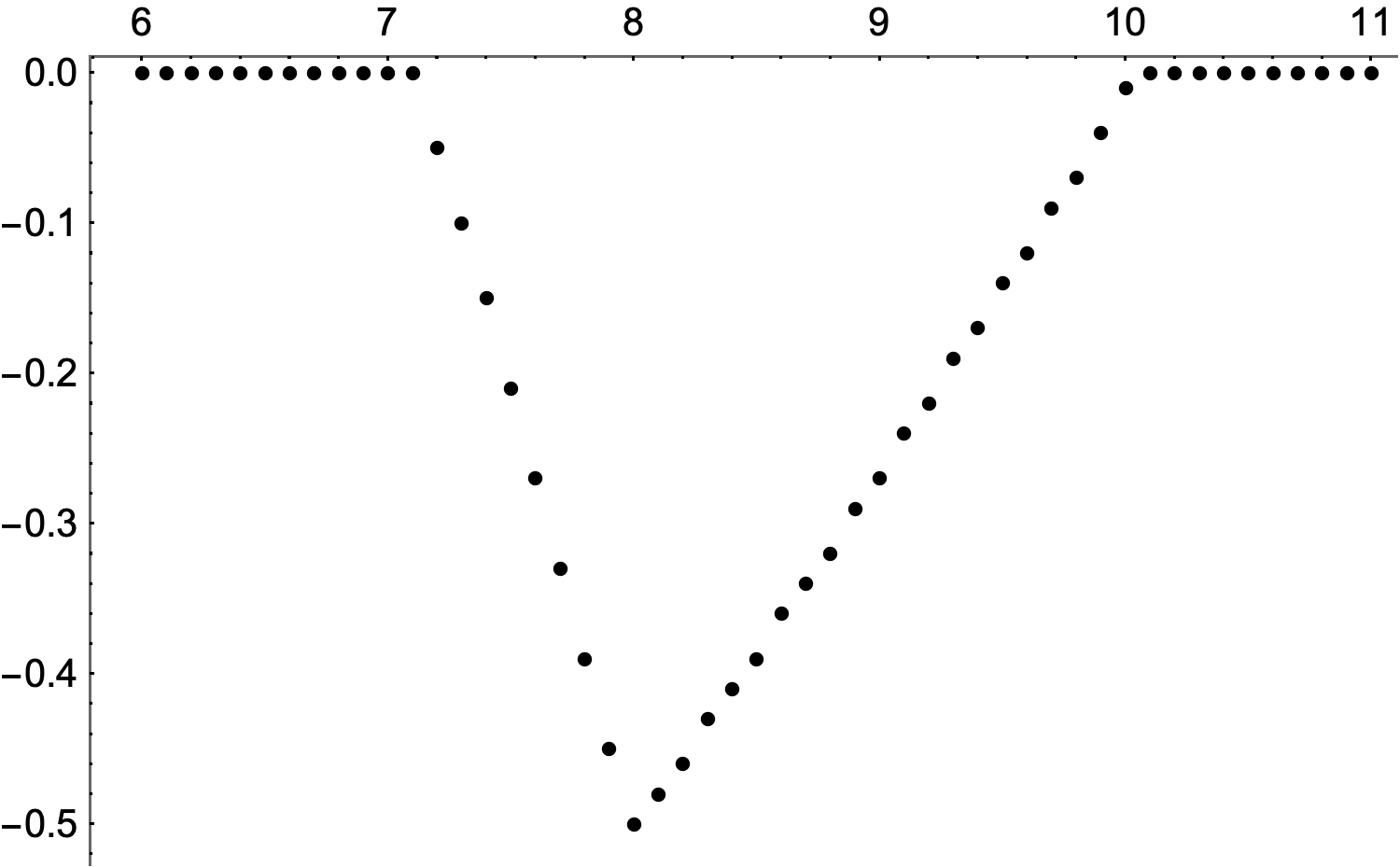}
\end{tabular}
\\
\hline
\begin{tabular}{l}
$\b = 0.25$
\end{tabular} &
\begin{tabular}{l}
\includegraphics[width=.25\linewidth]{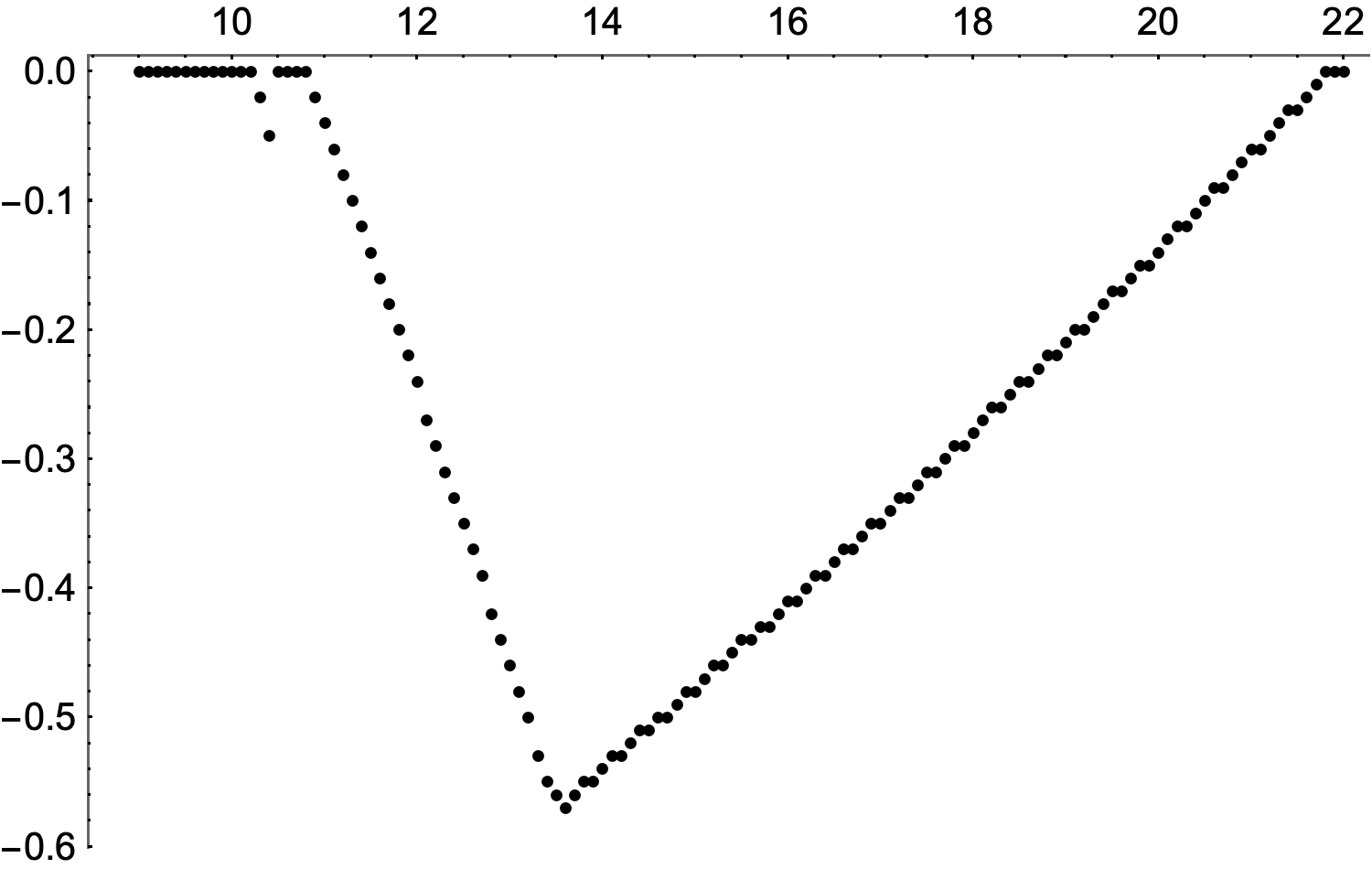}
\end{tabular} &
\begin{tabular}{l}
\includegraphics[width=.25\linewidth]{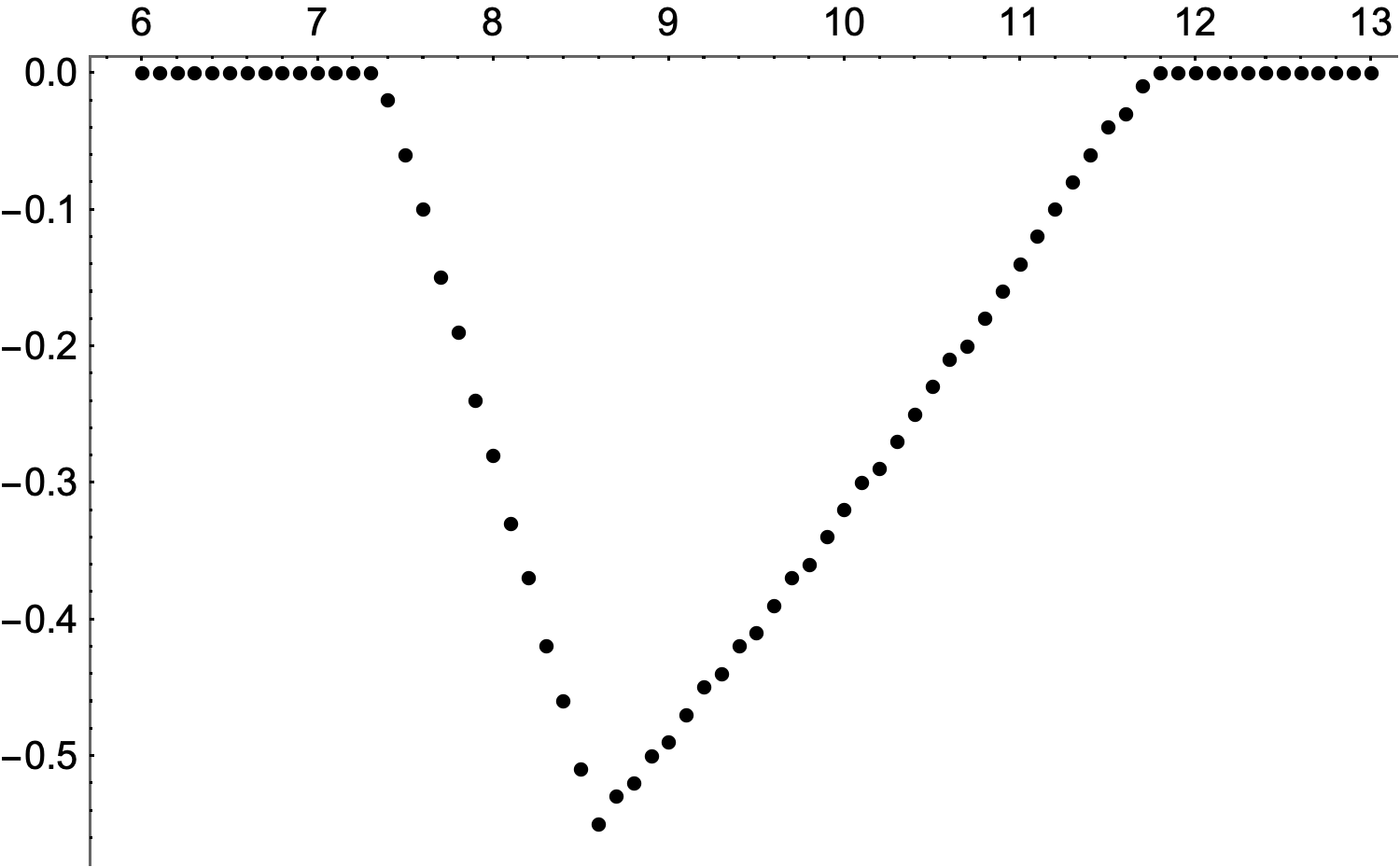}
 \end{tabular} 
&
\begin{tabular}{l}
\includegraphics[width=.25\linewidth]{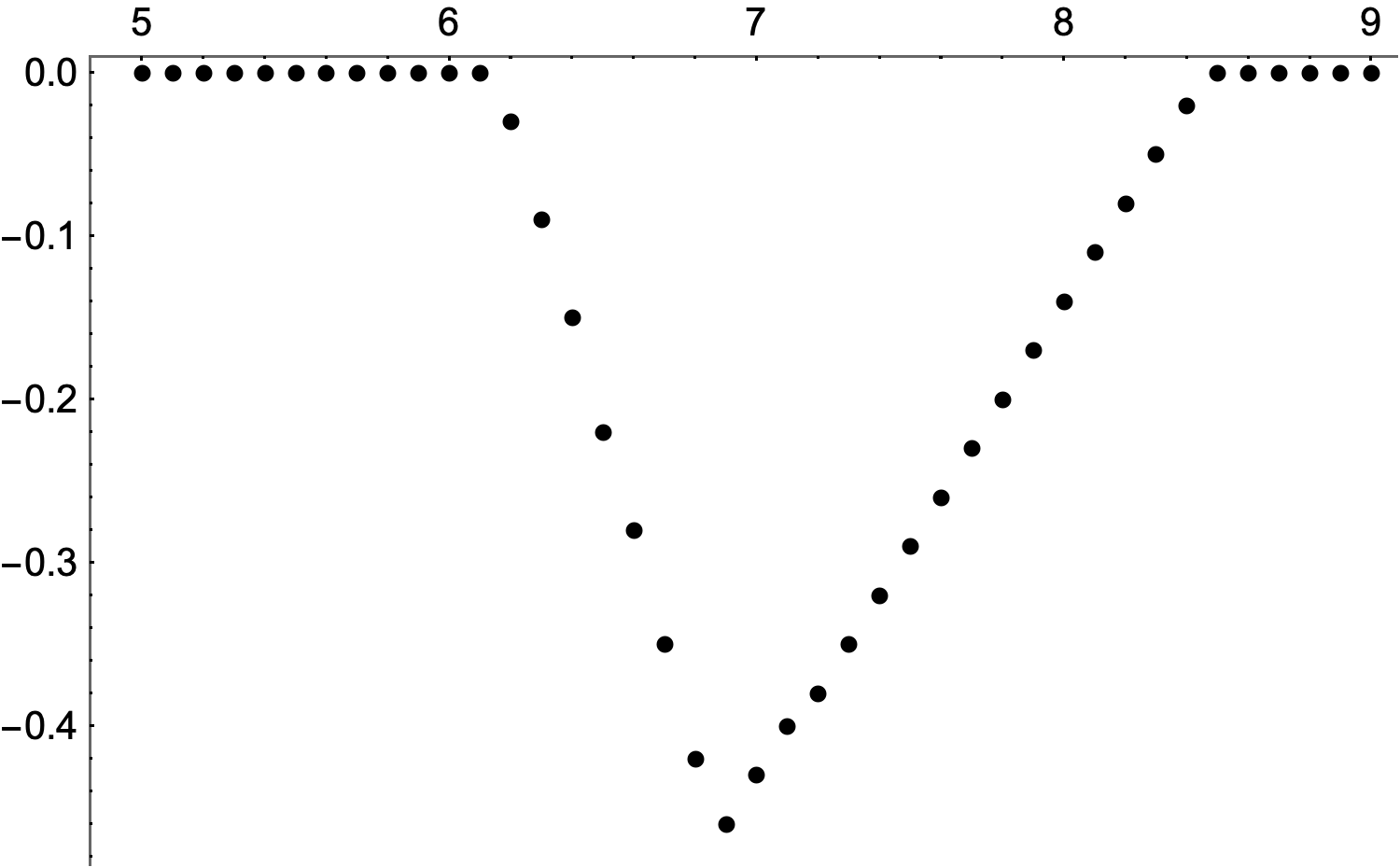}
\end{tabular}
\end{tabular}
\caption{Plots of $Q/M$ vs. $r_0/M$ for particles seeded on the $x$-axis with varying $\ol{q} B_0$ and $\b$. The red circled point and the orange boxed point in the upper right plot correspond to the trajectories in Figs.\ \ref{fig:regionII_trajectories} and\ \ref{fig:regionIII_trajectories}, respectively.}
    \label{fig:trianglefullgrid}
\end{figure}

We have seen above that the different behaviors of the charges in the four intervals summarized in Table\ \ref{tab:Bfieldsummary} ultimately arise from the regions shown in Fig.\ \ref{fig:Qxplot}. In our physically motivated simulation, where $Q_0 = 0$ and $Q$ is incremented as the particles fall in, only one portion of the $x_0$-$Q_0$ plane is probed, yielding the four intervals.

Having explained the origin of these behaviors, we now study the effect of changing the speed of the black hole and the strength of the magnetic field. We consider the following set up: We seed positive and negative charges at a distance $r_0$ on the positive $x$-axis. All the charges have an initial velocity of $\b$ in the positive $y$ direction, with initial conditions set as described above. We then increment or decrement the charge of the black hole if either the positive or negative particle (but not both) falls in and then we seed two more charges at the same spot. We repeat this process until the charge of the black hole stabilizes. We then increment the starting position and repeat the process. This generates a plot of the final charge of the black hole $Q$ as a function of the starting position $r_0$. We assume the mass of the black hole is effectively unperturbed for particles with a large charge and small mass. 

Consistent with Fig.\ \ref{fig:Qxplot}, we see triangular structures as shown in Fig.\ \ref{fig:trianglefullgrid} wherein the charge acquisition as a function of radius shows the behavior described above across the four intervals. We show nine such triangles for different values of $\b$ and $\ol{q} B_0$. The specific values of $r_I$, $r_{II},$ and $r_{III}$ depend on these parameters but the qualitative behavior is the same across Fig.\ \ref{fig:trianglefullgrid}. The trajectories shown in Figs.\ \ref{fig:regionII_trajectories} and\ \ref{fig:regionIII_trajectories} correspond to red and orange points highlighted in the upper right plot of Fig.\ \ref{fig:trianglefullgrid}. The behaviors shown in these figures are consistent along each leg of the triangle. As explained above, these triangles arise from the boundaries between the regions in Fig.\ \ref{fig:Qxplot}.

If instead we were to seed particles on the negative $x$-axis, the triangle would flip vertically around the $Q=0$ axis. If the distribution of charge is symmetric about $y$, then the net charge is zero, although charges will continue to flux around the black hole since $\mathbf{E}\cdot \mathbf{B}\ne 0$.

\begin{figure}[h]
    \centering
    \includegraphics[width=0.65\linewidth]{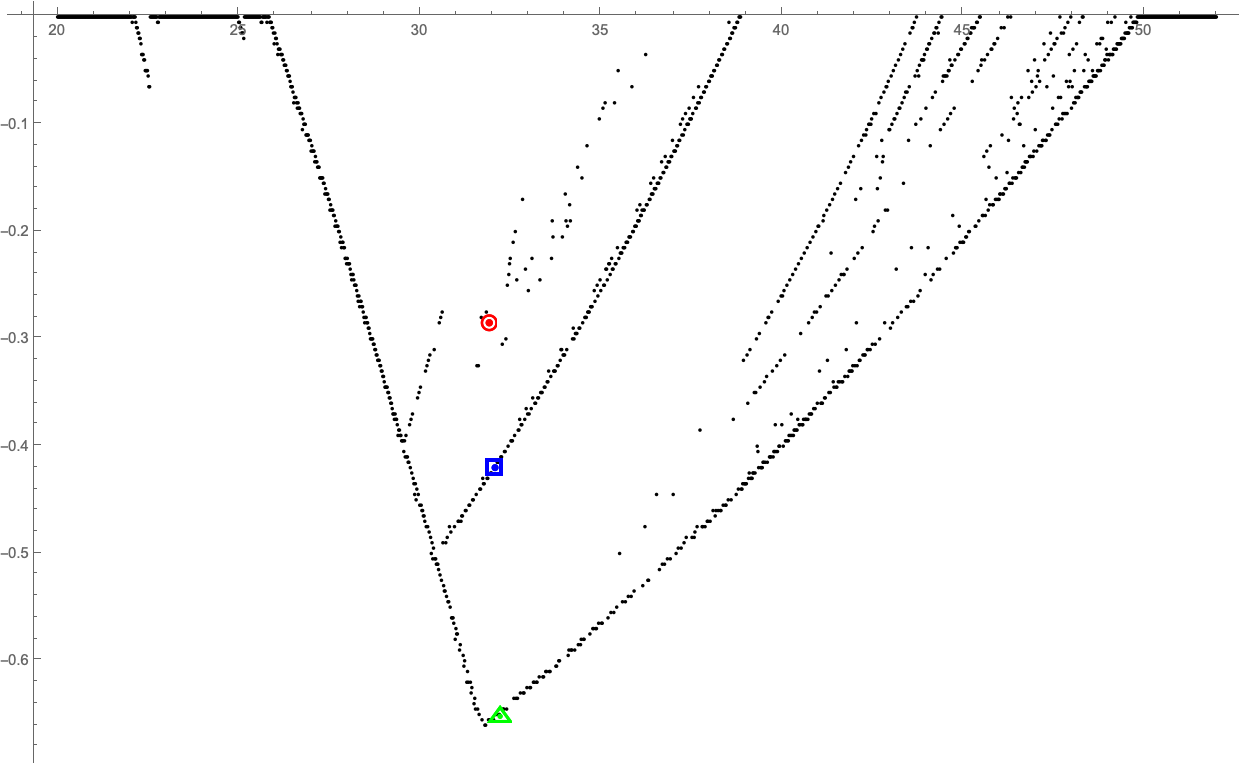}
    \caption{The triangle from the upper left of Fig.\ \ref{fig:trianglefullgrid} with higher resolution. The red circle, blue square, and green triangle correspond to the trajectories in Fig.\ \ref{fig:weirdplots}.}
    \label{fig:highres_triangle}
\end{figure}

\begin{figure}[htb]
\begin{tabular}{ccc}
\begin{tabular}{l}
\includegraphics[width=.3\linewidth]{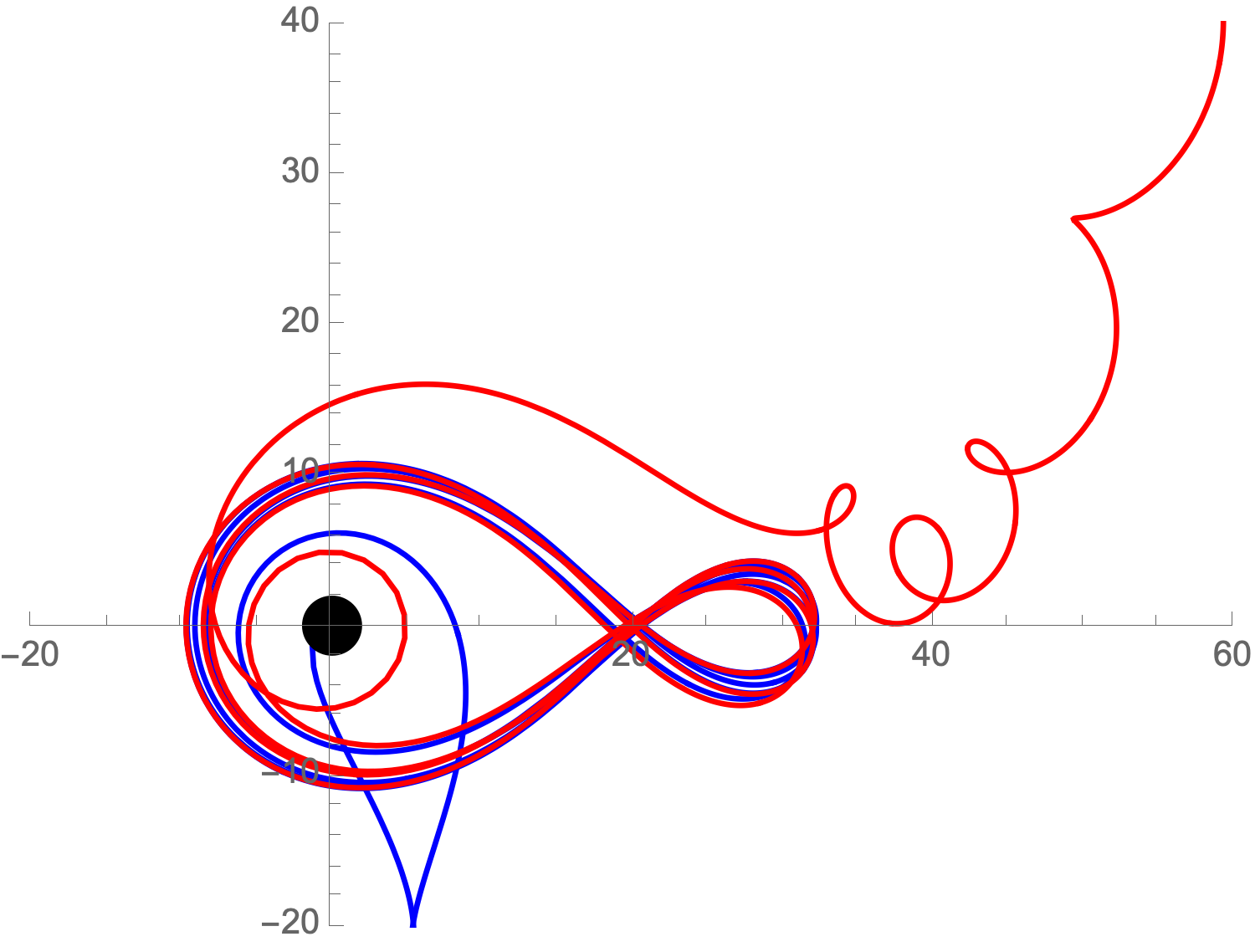}
\end{tabular} &
\begin{tabular}{l}
\includegraphics[width=.3\linewidth]{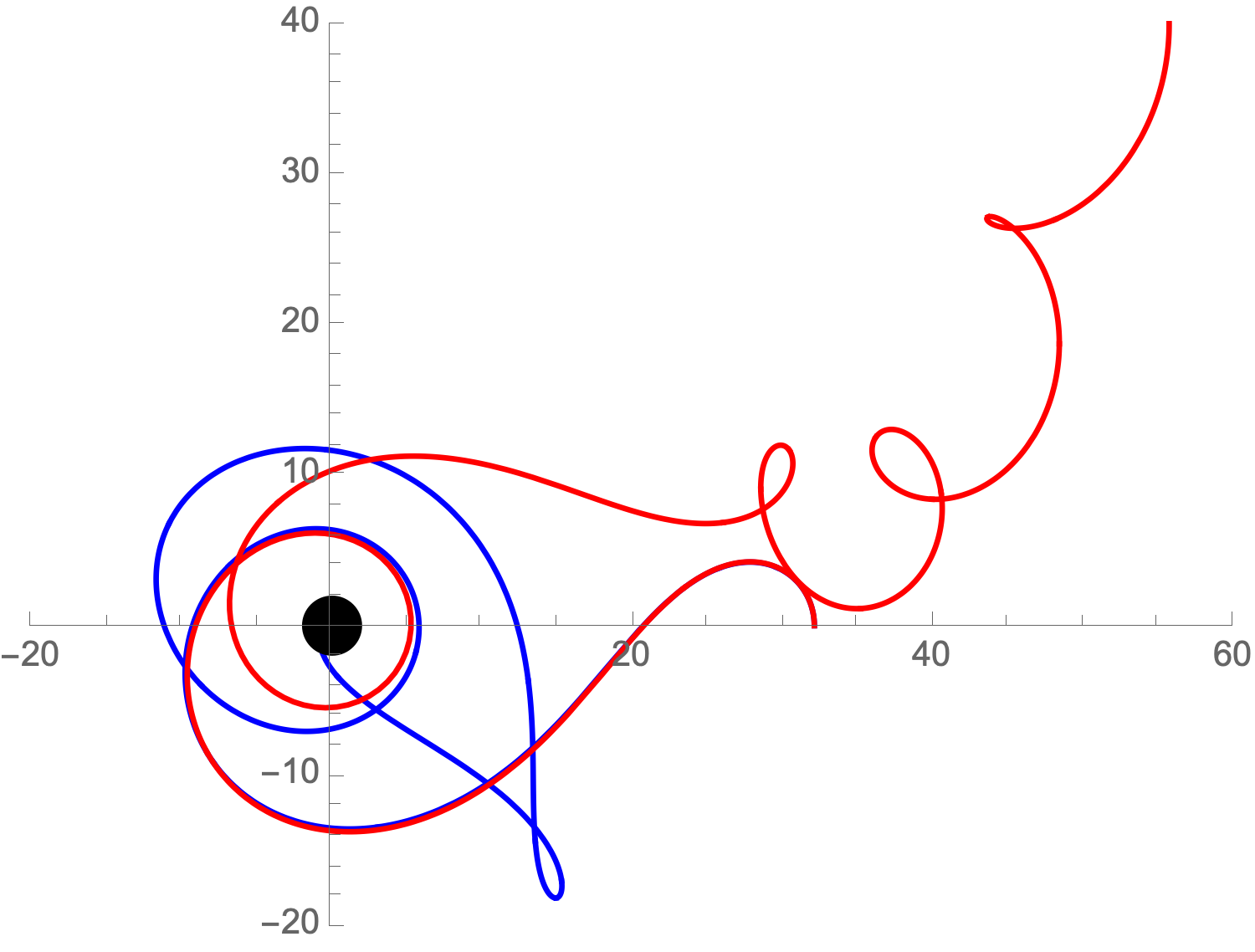}
 \end{tabular} &
\begin{tabular}{l}
\includegraphics[width=.3\linewidth]{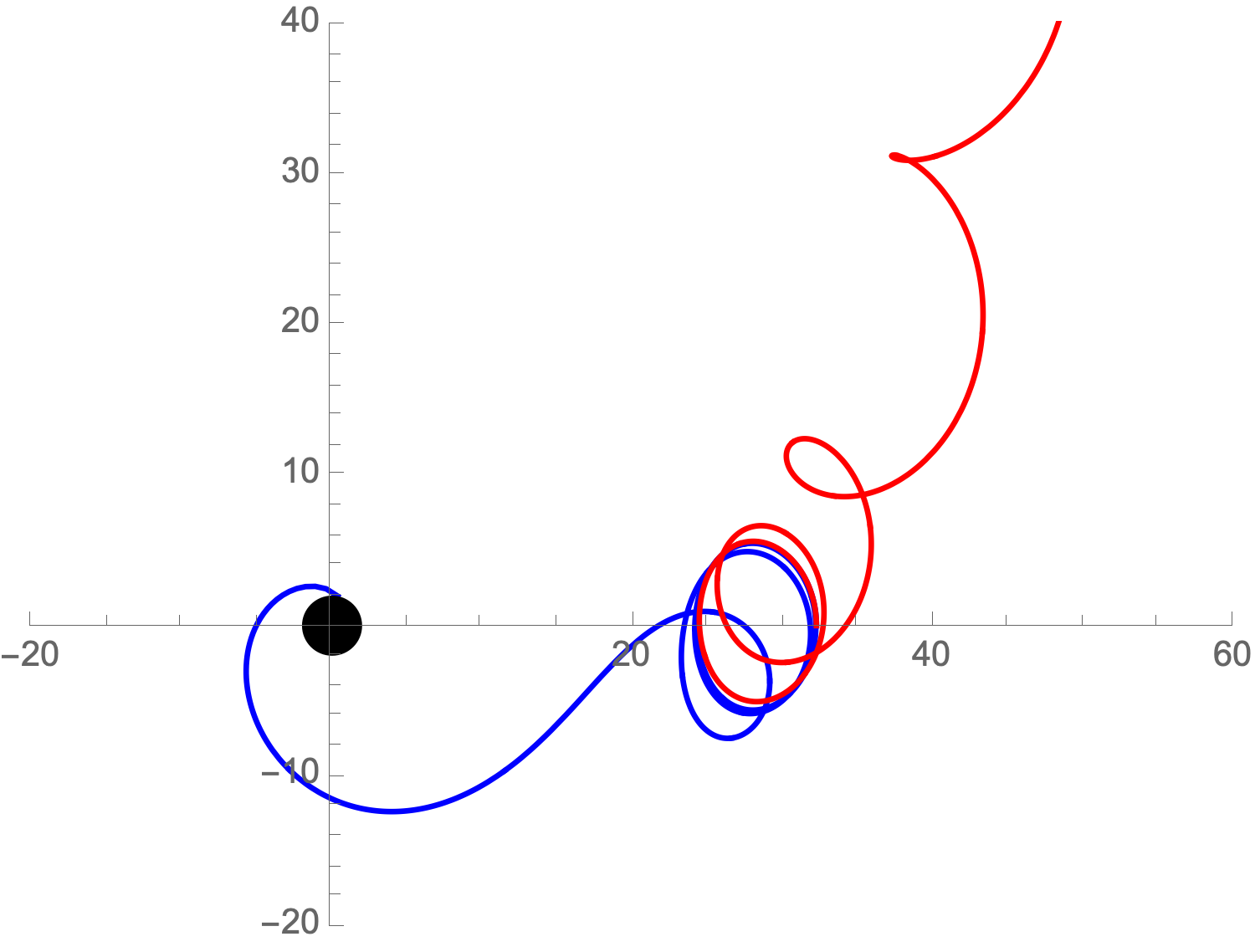}
\end{tabular}
\end{tabular}
\caption{Trajectories corresponding to the red circle, blue square, and green triangle in Fig.\ \ref{fig:highres_triangle}, respectively. All have $\ol{q}B_0 = 0.01 M^{-1}$ and $\b = 0.05$, and have $r_0 = 31.92M$, $32.08M$, and $32.22M$, respectively. All trajectories are for negative particles, and the two trajectories in the same plot correspond to slightly different values of $Q$, separated by increments of $\D Q = 0.005M$. The blue trajectory has the greater (less negative) $Q$, and the red trajectory's escape is what stabilizes the charge at its final value. In the first plot, the negative particles orbited the black hole many times in a figure-eight trajectory until falling in, or, for sufficiently negative $Q$, escaping. This stabilized the charge at $Q = -0.285M$, far above the points lying on the triangle. In the second plot, the two trajectories overlap until the charges approach the black hole, wherein one falls in and one escapes, stabilizing the charge at $Q = -0.420M$, still above the other points on the triangle but forming a new line in the plot. In the third plot, the usual behavior is observed, wherein the charges undergo many spirals before falling in or escaping, as in Fig.\ \ref{fig:regionIII_trajectories}. The final charge here was $Q = -0.650M$, along the leg of the triangle.}
\label{fig:weirdplots}
\end{figure}

The boundaries $r_I$, $r_{II}$, $r_{III}$ appear to be zero-dimensional on the $x$-axis for most of the triangles in Fig.\ \ref{fig:trianglefullgrid}. However, the top left triangle is broken up in a complex way. We probe three details of the broken triangle shown at higher resolution in Fig.\ \ref{fig:highres_triangle}. The three results indicated by a red circle, blue square, and green triangle are all very near each other in the space of initial conditions with values of $r_0$ within $0.3M$ of each other, yet with greatly disparate outcomes. The specific trajectories corresponding to these outcomes are shown in Fig.\ \ref{fig:weirdplots}. In each panel we are seeing in-falling trajectories of negative charges transition to escape trajectories at the corresponding value of $Q$ highlighted in Fig.\ \ref{fig:highres_triangle}. The extreme delicacy of the outcome is typical of chaotic systems. 

As we move into the rest of the equatorial plane, one might expect the boundaries to be one-dimensional, that is, lines that define the basin boundaries between the types of dynamical behavior. However, given that we anticipate chaotic behavior, we should rather expect these boundaries to be fractal \cite{Dettmann:1995ex, Levin:1998rk,  Levin:1999zx, Levin:2000md, Cornish:2002eh, Cornish:2003ig, Levin:2006zv}, reflecting the extreme sensitivity to initial conditions and the mixing and folding of trajectories that characterize chaos. The first triangle of Fig.\ \ref{fig:trianglefullgrid}, blown up in Fig.\ \ref{fig:highres_triangle}, is already demonstrating such behavior. Below we show that this is pervasive in the equatorial plane: the boundaries between zero charge and charge are fractal.

\subsection{The Equatorial Plane and Fractal Basin Boundaries}

\begin{figure}[tbp]
\begin{tabular}{>{\centering\arraybackslash} m{5em}|ccc}
& $|\ol{q}|B_0 = 0.01 M^{-1}$ & $|\ol{q}| B_0 = 0.05 M^{-1}$ & $|\ol{q}|B_0 = 0.1 M^{-1}$ \\ \hline
\begin{tabular}{l}
$\b = 0.05$
\end{tabular} &
\begin{tabular}{l}
\includegraphics[width=.25\linewidth]{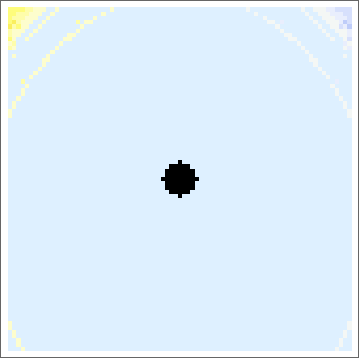}
\end{tabular} &
\begin{tabular}{l}
\includegraphics[width=.25\linewidth]{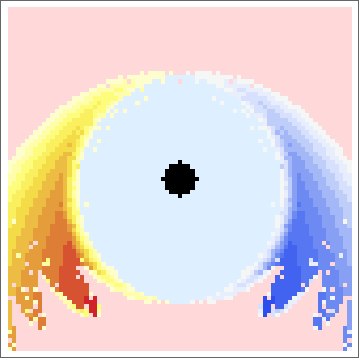}
 \end{tabular} &
\begin{tabular}{l}
\includegraphics[width=.25\linewidth]{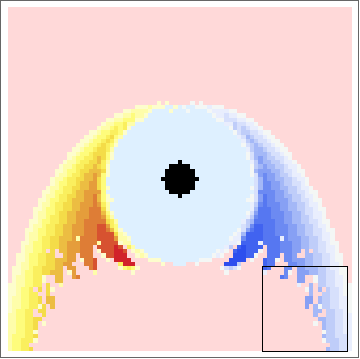}
\end{tabular}
\\
\begin{tabular}{l}
$\b = 0.15$
\end{tabular} &
\begin{tabular}{l}
\includegraphics[width=.25\linewidth]{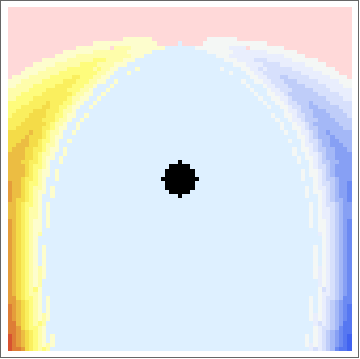}
\end{tabular} &
\begin{tabular}{l}
\includegraphics[width=.25\linewidth]{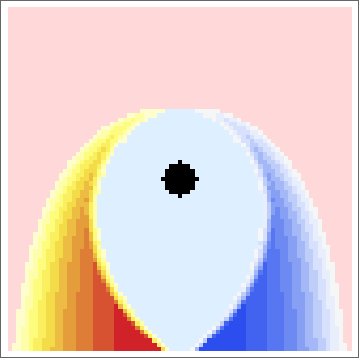}
 \end{tabular} &
\begin{tabular}{l}
\includegraphics[width=.25\linewidth]{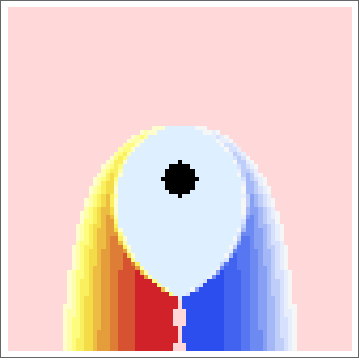}
\end{tabular}
\\
\begin{tabular}{l}
$\b = 0.25$
\end{tabular} &
\begin{tabular}{l}
\includegraphics[width=.25\linewidth]{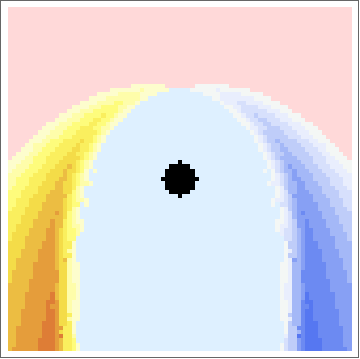}
\end{tabular} &
\begin{tabular}{l}
\includegraphics[width=.25\linewidth]{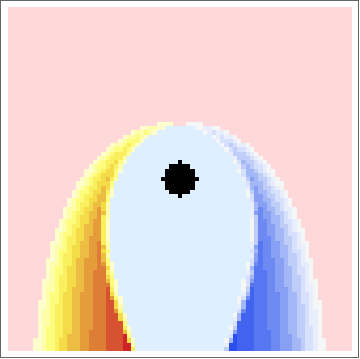}
 \end{tabular} 
&
\begin{tabular}{l}
\includegraphics[width=.25\linewidth]{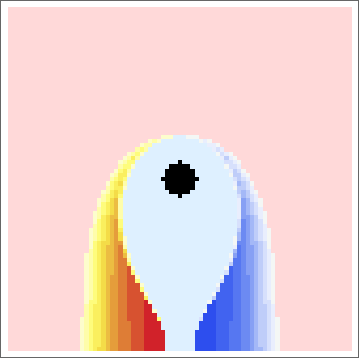}
\end{tabular}\\
& \multicolumn{3}{c}{\includegraphics[width=.9\linewidth]{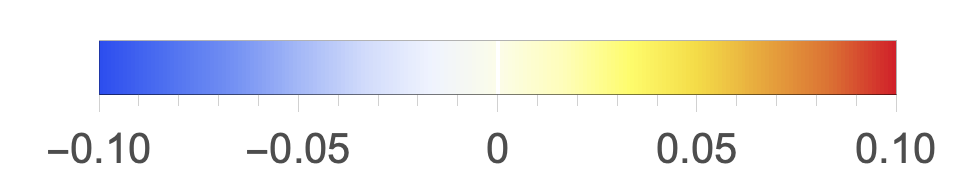}}
\end{tabular}
\caption{Plots of $Q$ in the $x$-$y$ plane for varying $|\ol{q}|B_0$ and $\b$. All plots have $-20M \leq x, y \leq 20 M$. If both positive and negative particles seeded at the $x$ and $y$ for that cell fall into the black hole, then the cell is colored light blue. If both particles escape, the cell is colored light red. If one falls in and one escapes, the charge of the black hole is increased or decreased until it stabilizes. The spectrum from dark blue to dark red represents the final charge in units of $M$. The range of the triangles on the $x$-axis are visible in most of these panels, although the restricted standardized range of these figures doesn't cover the same ranges as those in Fig.\ \ref{fig:trianglefullgrid}. The box in the corner of the upper right plot corresponds to the region examined in the first plot of Fig.\ \ref{fig:fractal_plots}.}
\label{fig:2d_chargegrid}
\end{figure}

Extending our analysis to the entire equatorial plane, the triangles are represented in the basin boundaries of Fig.\ \ref{fig:2d_chargegrid}. The simulation proceeds much as in the previous section. We divide the region $-20M \leq x, y \leq 20 M$ into cells of size $0.1M$, seed positive and negative particles at the top left corner of each cell, and color the cell based on the behavior of the charges. As before, if one particle falls in and the other escapes, the black hole's charge is increased or decreased and the cell is seeded again, repeating until the charge has stabilized. The cell is then colored to represent the final charge of the black hole as shown in the legend. Extracting the data from the positive $x$ axis alone would give Fig.\ \ref{fig:trianglefullgrid}. Reversing the signs of $x_0$, $\ol{q}$, and $Q$ reflects the force about the $y$-axis, leading to the antisymmetry about $y$ in the figures. Again, charge distributions that are symmetric about the $y$-axis yield $Q=0$, asymmetric distributions yield $Q \neq 0$, and in both cases, charged particles continue to flux since $\mathbf{E}\cdot\mathbf{B}\ne 0$.

As anticipated, the boundaries show evidence of chaotic behavior. For a dynamical system with $N$ coordinates and $N$ canonical momenta, the motion in phase space will be restricted to tori if there are $N$ constants of motion. The argument is simply that a set of coordinates can always be found such that each canonical momentum is set equal to a constant, creating a set of constant frequencies. The corresponding coordinates are then the angular variables on the $N$-torus. All integrable systems, that is, non-chaotic systems, have this feature.

As constants of motion are lost, the system becomes non-integrable with no closed-form analytic solution as a function of initial conditions. The system is still deterministic, however the loss of constants of motion allows orbits to exist off of torii in phase space. The orbits show extreme sensitivity to initial conditions and a mixing and folding of trajectories, as expected with chaotic systems.

A spinning black hole has four constants of motion: energy, angular momentum about the spin axis, $u\cdot u$, and the Carter constant. Therefore the orbits are integrable. In boosting the black hole through the magnetic field, we have lost constants of motion. We expect the most general orbits to be non-integrable and chaotic. A nice demonstration of chaotic behavior is the presence of fractal basin boundaries \cite{Dettmann:1995ex, Levin:1998rk, Levin:1999zx, Levin:2000md, Cornish:2002eh, Cornish:2003ig, Levin:2006zv}, which we happened to be scanning in our generalization to the equatorial plane of the triangles seen in one dimension. 

\begin{figure}[tb]
    \centering
        \begin{tabular}{ccc}
        \includegraphics[width=.32\linewidth]{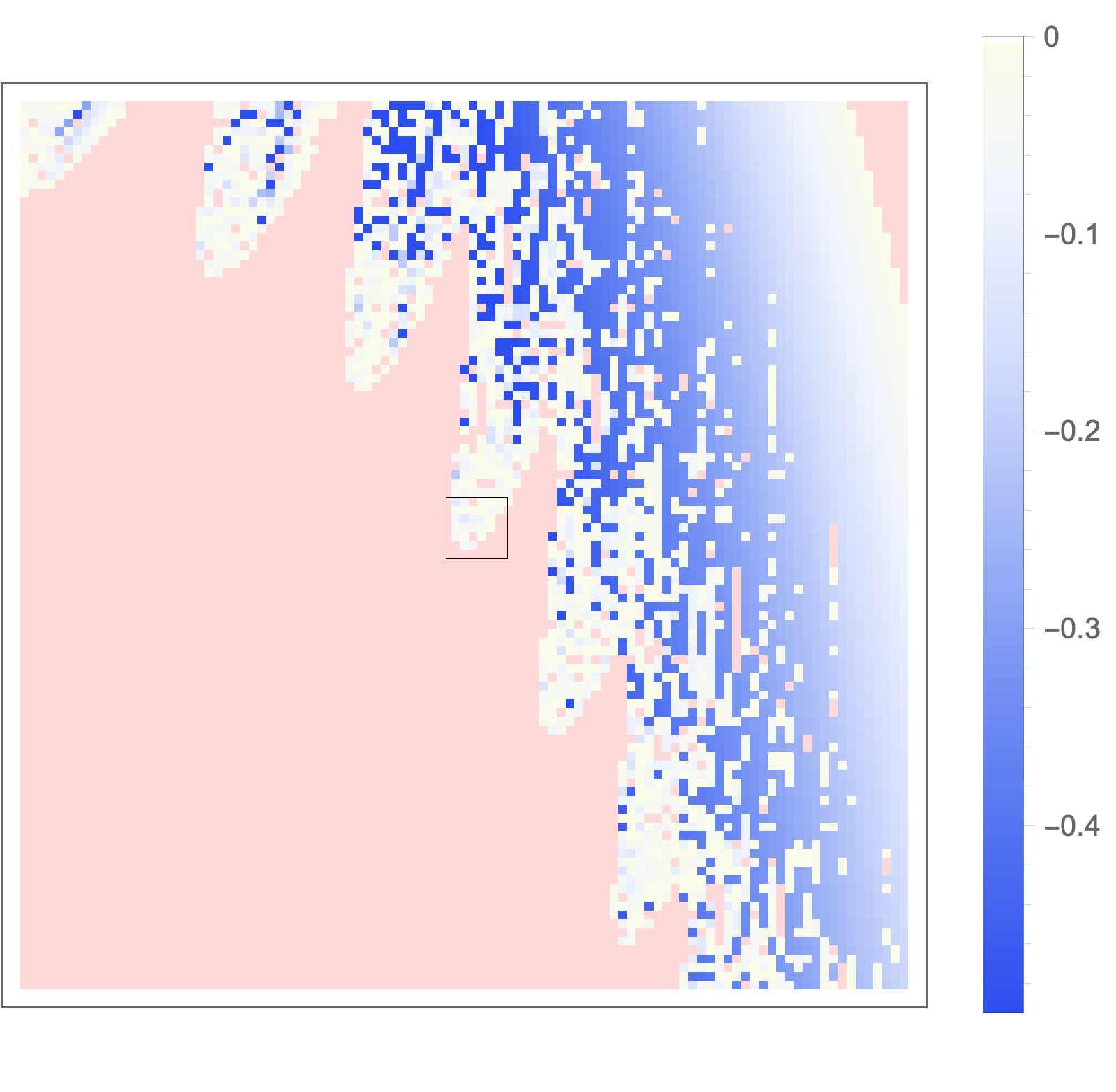} &  \includegraphics[width=.32\linewidth]{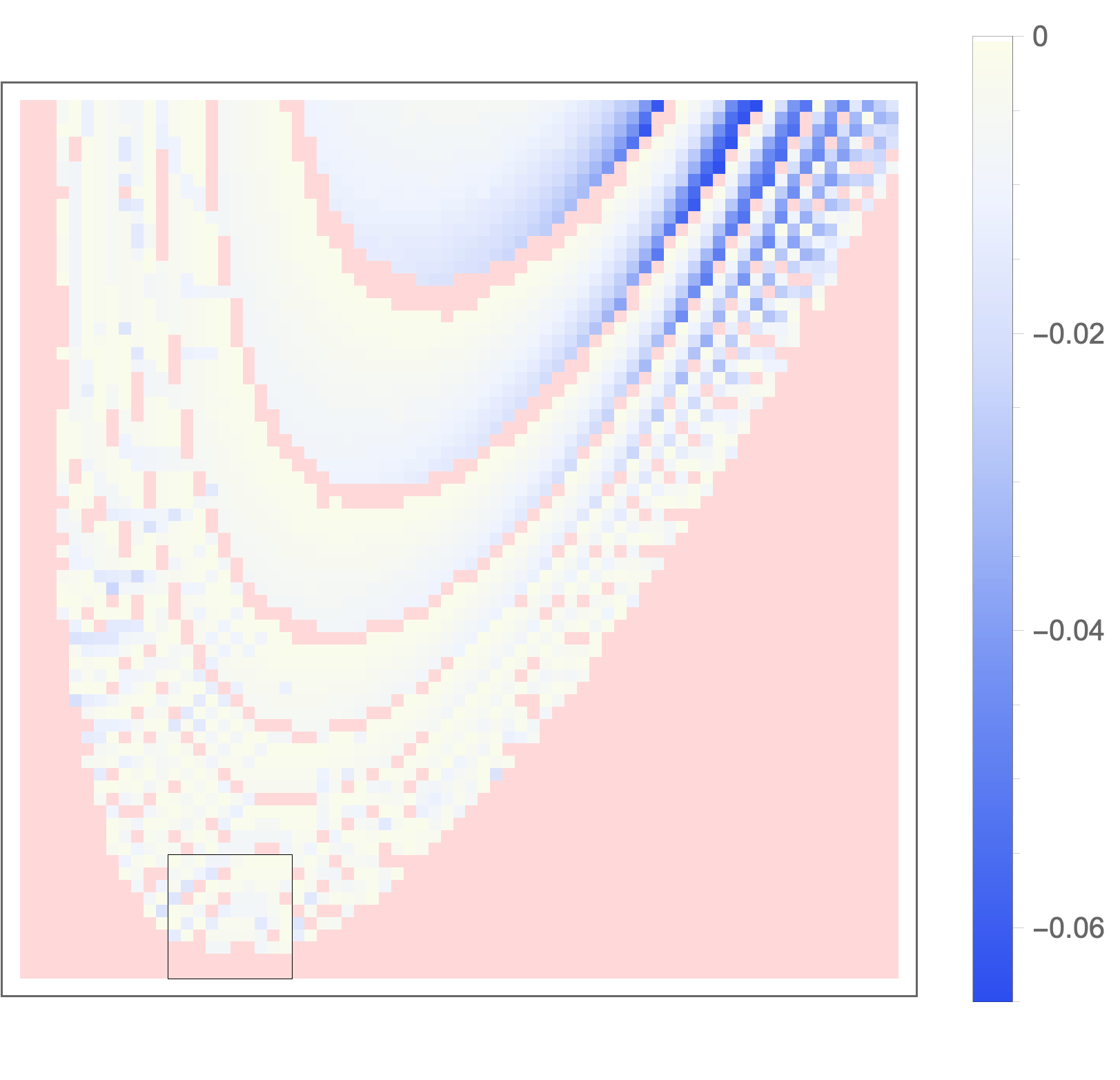} &
        \includegraphics[width=.32\linewidth]{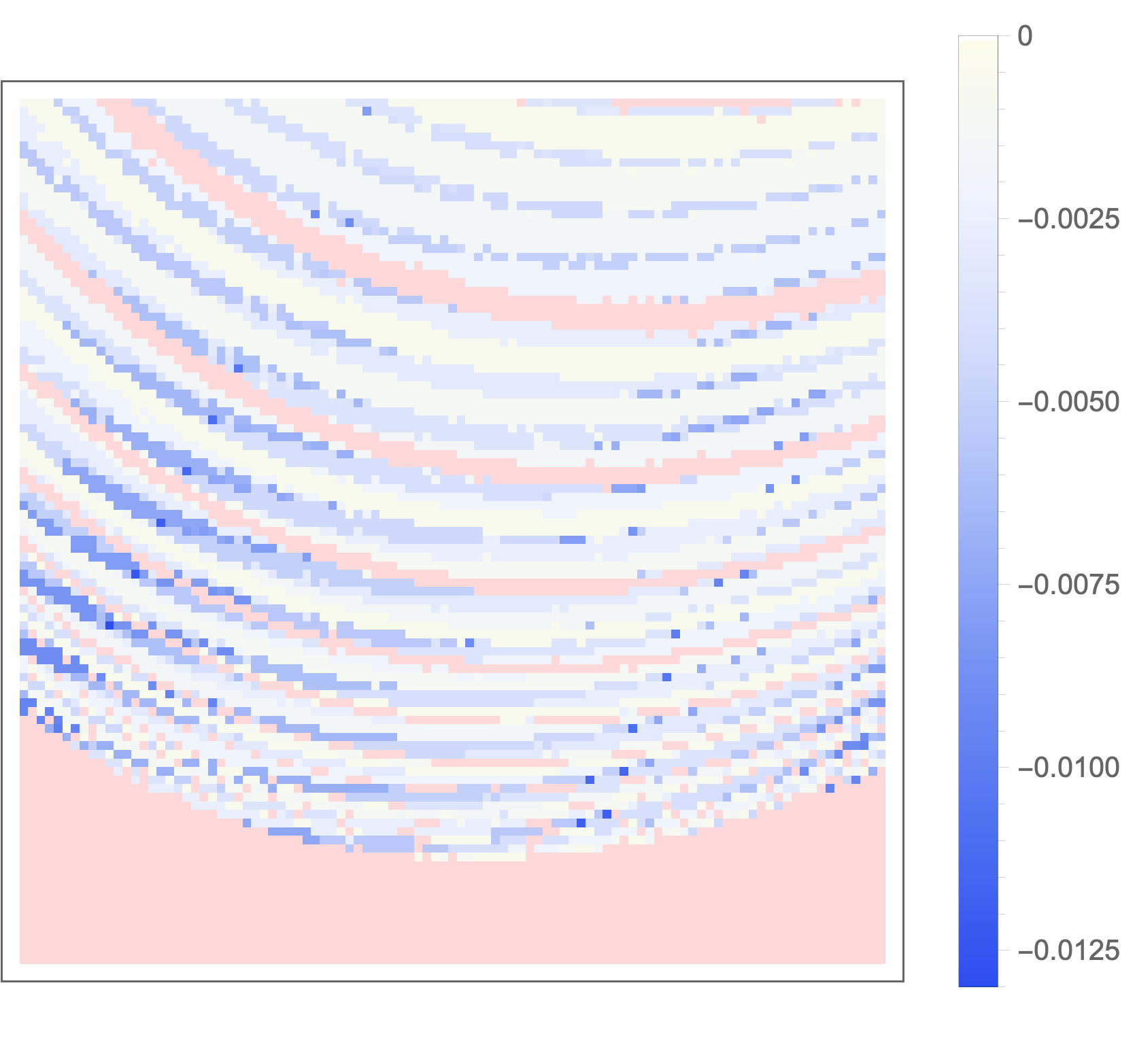} \\
    \end{tabular}
    \caption{Basin boundaries at various scales. All have $\ol{q}B_0 = 0.1 M^{-1}$ and $\b = 0.05$. The first plot has $10 M \leq x \leq 20 M$, $-20 M \leq y \leq -10 M$, and a spacing of $0.1M$. The second has $14.84 M \leq x \leq 15.54 M$, $-15.10 M \leq y \leq -14.40 M$, and a spacing of $0.01M$. The third has $14.96 M \leq x \leq 15.06 M$, $-15.10 M \leq y \leq -15.00 M$, and a spacing of $0.001M$.}
    \label{fig:fractal_plots}
\end{figure}

The substructure is already apparent in Fig.\ \ref{fig:2d_chargegrid}. To see that the fractal structure persists, we zoom in on the region shown in the box in the upper right plot in Fig.\ \ref{fig:2d_chargegrid}. The results are show in Fig.\ \ref{fig:fractal_plots}, where we continue to zoom in on smaller regions. The repetition of structure on smaller and smaller scales is the definition of a fractal. The mixing and folding of trajectories is evident in the mixed colors at the boundary. And the sensitivity to initial conditions is evident as we zoom in and still see that a minuscule deviation in the initial data leads to very different orbits.

\begin{figure}[!hbt]
  \centering
  \includegraphics[width=.65\linewidth]{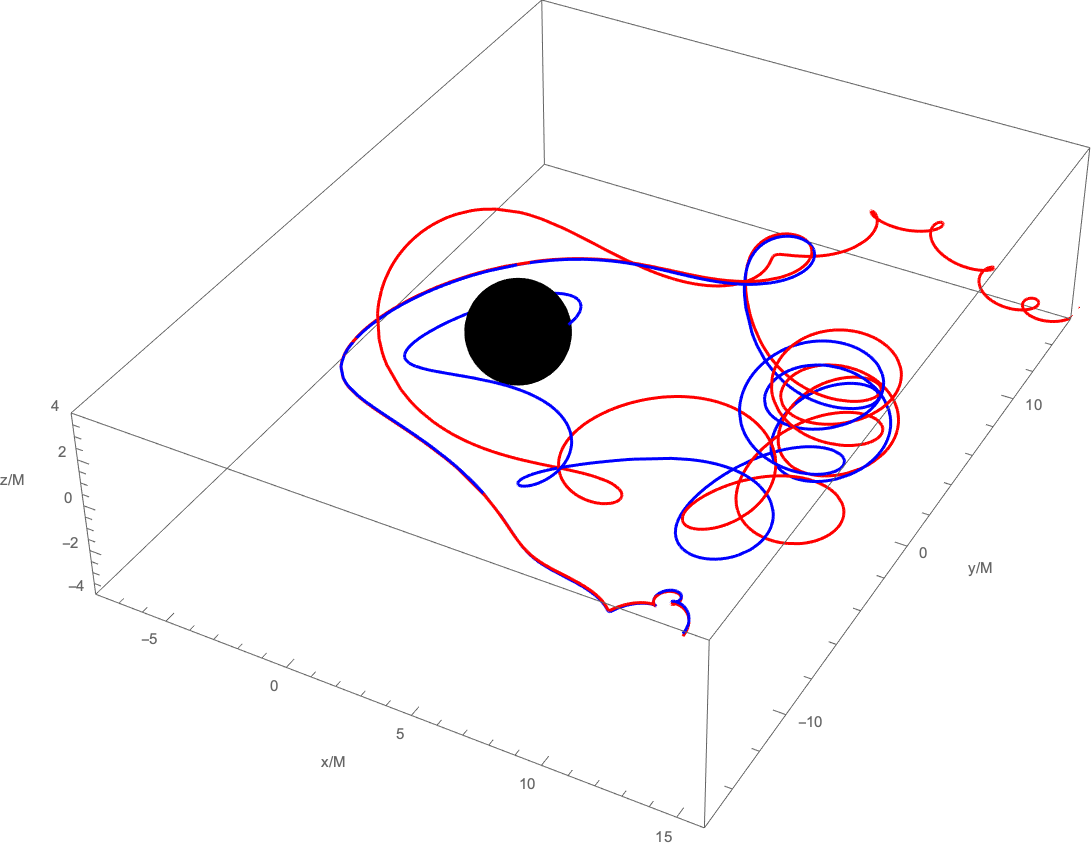}
  \caption{A pair of non-equatorial trajectories. Both particles are negative, and have $x_0 = 14.6M$ and $y_0 = -15M$. The red trajectory has $z_0 = 2.8M$, and the blue trajectory has $z_0 = 2.9 M$. The system parameters are $\ol{q} B_0 = 0.1 M^{-1}$, $\b = 0.1$, and $Q = 0$.}
  \label{fig:3D}
\end{figure}

To close this section, we show a pair of nearby non-equatorial trajectories in Fig.\ \ref{fig:3D}. As is evident, the small difference in initial conditions yields very different behavior, as one particle escapes and one ends in the black hole. This demonstrates that the chaotic behavior exists outside the equatorial plane as well, and we expect fractal basin boundaries to appear throughout space, though investigating the entire dynamics is beyond the scope of this paper.

\section{Summary}

As anticipated, a black hole boosted through an ambient magnetic field can preferentially acquire charge. The charge acquired depends on the details of the initial data and the distribution of particles around the black hole. Spatially symmetric distributions lead to $Q=0$; spatially asymmetric distributions to black hole charge. Regardless, charged particles continue to flux around the black hole. Our analysis is not in any way intended to be a model of a realistic astrophysical set up. We simply intend to demonstrate the general principle: All motion through magnetic fields will contribute to the charge of a black hole in different ways. Another general conclusion is that the gravito-electrodynamics determine the actual outcome of charge accretion, not the screening of the electric field. In our case, the gravito-electrodynamics are chaotic as evidenced by the fractal basin boundaries between outcomes.

Once charged, the black hole can support its own electromagnetic field leading to its own intrinsic luminosity. Possible astrophysical settings for boosted black holes include supermassive black holes that are boosted relative to the central galactic magnetic field, black holes with magnetized and offset accretion disks, and black holes orbiting through the magnetosphere of a companion neutron star. The observability of these effects could be assessed using something like the Larmor frequency. However, given the chaotic dynamics, and the consequent extreme sensitivity on initial data, there are no general conclusions to draw from that frequency for any specific orbit. A power estimate such as the circuit paradigm of \cite{McWilliams:2011zi} for a related charged system suggests detection will be challenging but not impossible. Still, we are excited about the prospects for observing charged bare black holes during this prolific black hole century through surveys, the EHT project, and the multi-messenger networks around LIGO-VIRGO-KAGRA.

\section*{Acknowledgements}

We thank Albert Law and Kshitij Gupta for collaborative conversations. RB is supported in part by the U.S. Department of Energy grant DE-SC0011941 and Simons Foundation Award Number 555117. JL is supported in part by the Tow Foundation. 

\appendix
\section{Solving the Maxwell Equations}

The Maxwell equations are given in \eqref{Maxwell}, but we repeat them here:
\be
\del_\n\left(\sqrt{-g}F^{\m\n}\right) = 0, \quad
\del_\n\left(\sqrt{-g} (\star F)^{\m\n}\right) = 0.
\ee
Following \cite{Morozova:2013ina}, we take the nonvanishing components of the vector potential to be $A_t\left(r,\th'\right)$ and $A_\f(r,\th)$, where $\th'$ is the ``polar'' angle down from the positive $x$ axis. We have chosen this unusual angle because we have an electric field aligned with the negative $x$ axis at infinity. With this assumption, the dual Maxwell equations are automatically satisfied. 

Taking $\m = t$ and $\m = \f$ in the first equation yields
\begin{align}
N^2 \frac{\del}{\del r}
\left(r^2 \frac{\del A_t}{\del r}\right)
+ \frac{1}{\sin\th'} \frac{\del}{\del \th'}
\left(\sin\th' \frac{\del A_t}{\del \th'}\right) 
&= 0\\
r^2 \frac{\del}{\del r}
\left(N^2 \frac{\del A_\f}{\del r}\right)
+ \sin\th \frac{\del}{\del \th}
\left(\frac{1}{\sin\th} \frac{\del A_\f}{\del \th}\right) &= 0,
\end{align}
where as before, $N^2 = 1-\frac{2M}{r}.$ The electric field is uniform and in the negative $x$ direction at infinity. In spherical coordinates, this becomes the boundary condition that for large $r$,
\begin{equation}
\left(E^r, E^\th, E^\f\right) \to 
-E_0\left(\sin\th \cos\f, \frac{1}{r}\cos\th\cos\f, -\frac{1}{r}\frac{\sin\f}{\sin\th}\right).
\end{equation}
Note that $\cos\th' = \sin\th\cos\f$. This gives the boundary condition for large $r$
\begin{equation}
A_t \to -E_0 r \cos\th'.
\end{equation}
We shall also impose that $A_t$ vanish at the horizon. Since we also have a uniform magnetic field in the $z$ direction at infinity, we also have the boundary condition 
\begin{equation}
\left(B^r, B^\th, B^\f\right) \to 
B_0\left(\cos\th \cos\f, -\frac{1}{r}\sin\th, 0\right).
\end{equation}
This implies
\begin{equation}
A_\f \to \frac{1}{2} B_0 r^2 \sin^2\th.
\end{equation}
We also impose that $A_\f$ vanish at the origin.

There is one additional boundary condition that we impose, that of regularity of the observed energy density $\r_\text{obs} = T_{\m\n} u^\m u^\n$ at the horizon, where \begin{equation}
u^\m = \left(\left(1-\frac{2M}{r}\right)^{-1}, -\sqrt{\frac{2M}{r}}, 0, 0\right)^\m
\end{equation}
is the four-velocity of a freely falling observer. It can be shown that this forces $A_t$ to vanish at the horizon at least linearly, while $A_\f$ simply cannot diverge at the horizon. 

We now take $A_t = f(r) P\left(\th'\right)$ in the first equation and separate variables, yielding
\begin{align}
\left(1 - \frac{2M}{r}\right)
\left(2 rf'(r) + r^2 f''(r)\right) - \ell(\ell+1)f(r) &= 0\\
\frac{1}{\sin\th'} \frac{d}{d\th'}
\left(\sin\th'
\frac{dP\left(\th'\right)}{d\th'}\right)
+ \ell(\ell+1)P\left(\th'\right) 
&= 0.
\end{align}
We have written the separation constant as $\ell(\ell+1)$ for later convenience. We shall take $\ell$ to be real, as complex $\ell$ yields non-regular solutions. The angular equation is Legendre's differential equation, solved by the Legendre functions of the first and second kind, i.e., $P_\ell(\cos\th')$, and $Q_\ell(\cos\th')$. The Legendre functions of the second kind are singular at $\th = 0$ and $\p$, as are the Legendre functions of the first kind for non-integer $\ell$, so we take $P(\th) = P_\ell(\cos\th')$ and restrict to integer $\ell$. We note that the separation constant has the same value for $\ell = -n$ and $\ell = n-1$ for non-negative $n$, so we can take $\ell \geq 0$ as well.

We can rewrite the radial equation as
\begin{equation}
x(x-1)f''(x) + 2(x-1)f'(x) - \ell(\ell+1)f(x) = 0,
\end{equation}
where $x = r/2M$. This is a hypergeometric differential equation with parameters $a = -\ell$, $b = \ell+1$, and $c = 2$, so one solution is
\begin{equation}
f_\ell^{(1)}(x) = {}_2F_1(-\ell, \ell+1, 2; x).
\end{equation}
We note that this is a polynomial of degree $\ell$ since $\ell$ is a non-negative integer. It also vanishes at $x = 1$. The second solution, which can be found using the method of Frobenius, has a logarithmic contribution that diverges at the horizon for $\ell > 0$, so it can be ignored. We shall separately handle the $\ell = 0$ case, where the solutions are $1$ and $1/x$. We thus have
\begin{equation}
A_t\left(r,\th'\right) = 
C_0 + D_0\frac{2M}{r} + 
\sum_{\ell = 1}^\infty C_\ell f_\ell^{(1)}\left(r/2M\right)
P_m\left(\cos\th'\right)
\end{equation}
with arbitrary constants $C_0$, $D_0$, and $C_\ell$. To make $A_t$ vanish at the horizon, we impose the condition $C_0 + D_0 = 0$. Matching with $-E_0 r\cos\th$ at infinity yields $C_1 = 2 M E_0$ and $C_\ell = 0$ for $\ell > 1$. Thus our solution becomes
\begin{equation}
A_t\left(r,\th'\right) = 
C_0\left(1-\frac{2M}{r}\right) -
E_0 \left(1-\frac{2M}{r}\right) r \cos\th'.
\end{equation}
It turns out that the coefficient of the $1/r$ piece will contribute to the charge of the black hole. Since this black hole is uncharged, we set $C_0 = 0$, so we have
\begin{equation}
A_t\left(r,\th'\right) = -E_0 N^2 r \cos\th',
\quad A^t = E_0 r \cos\th'.
\end{equation} 
Moving on to the $A_\f$ equation, we take $A_\f = g(r) Q(\th)$ and separate variables, finding
\begin{align}
\left(1-\frac{2M}{r}\right)r^2 g''(r) 
+ 2M g'(r) - \ell(\ell+1)g(r) &= 0\\
\sin\th \frac{d}{d\th}
\left(\frac{1}{\sin\th}\frac{dQ(\th)}{d\th}\right)
+ \ell(\ell+1)Q(\th) &= 0.
\end{align} 
We can rewrite the radial equation as
\begin{equation}
x(x-1)g''(x) + g'(x) - \ell(\ell+1)g(x) = 0,
\end{equation}
where again $x = r/2M$. This can be transformed into another hypergeometric by taking $g(x) = x^2 h(x)$. The first solution is
\begin{equation}
g_\ell^{(1)}(x) 
= x^2 \, {}_2F_1(\ell+2, 1-\ell, 3; x).
\end{equation}
This only has integers powers of $x$ in its Taylor series if $\ell$ is an integer. Since this must eventually be matched to $r^2$ for large $r$, we thus take $\ell$ to be an integer, in which case $g_\ell^{(1)}(x)$ goes as $x^{\ell+1}$ for large $x$. Again, we can take $\ell \geq 0$. Lastly, for $\ell = 0$, the first solution has a $\ln(1-x)$ contribution that diverges at the horizon, so we can ignore it. The second solution, as before, involves a logarithmic contribution for $\ell \neq 0$ that diverges at the horizon and can thus be ignored. For $\ell = 0$, the second solution is a constant and must be included.

The angular equation can be rewritten in terms of $u = \cos\th$:
\begin{equation}
(1-u^2)Q''(u) + \ell(\ell+1) Q(u) = 0.
\end{equation}
This can also be transformed into a hypergeometric equation by taking $z = u^2$.
The two solutions are
\begin{equation}
Q_\ell^{(1)}(u) = {}_2F_1\left(\frac{\ell}{2}, -\frac{\ell+1}{2}, \frac{1}{2}, u^2\right) 
\quad \text{ and } \quad
Q_\ell^{(2)}(u) = u\, {}_2F_1\left(\frac{\ell+1}{2}, -\frac{\ell}{2}, \frac{3}{2}, u^2\right).
\end{equation}
The second solution is not smooth at $\th = 0$ or $\p$ for $\ell > 0$, so we can ignore it for those values of $\ell$. For $\ell = 0$, the second solution is simply $u$. We thus have
\begin{equation}
A_\f(r,\th) = 
G_0 + G'_0 \cos\th
+ \sum_{\ell=1}^\infty F_\ell g_\ell^{(1)}\left(r/2M\right) Q_\ell^{(1)}\left(\cos^2\th\right)
\end{equation}
with arbitrary constants $G_0$, $G_0'$, and $F_\ell$. Clearly $g_\ell^{(1)}(0) = 0$, so we require $G_0 = G_0' = 0$ to make $A_\f$ vanish at the origin. Since $g_\ell^{(1)}(r)$ goes as $r^{\ell+1}$ for large $r$, the boundary condition at infinity sets $F_1 = 2 M^2 B_0$ and $F_\ell = 0$ for $\ell > 1$. Thus we have
\begin{equation}
A_\f = \frac{1}{2} B_0 r^2 \sin^2\th, \quad
A^\f = \frac{1}{2} B_0.
\end{equation}

Since the Maxwell equations are
linear, we can incorporate a charged black hole by simply adding the potential for a point charge, finding
\begin{align*}
A_t\left(r,\th'\right) 
&= -\frac{Q}{r} 
- E_0\left(1 - \frac{2M}{r} \right)
r \sin\th\cos\phi\\
A_\f(r,\th) 
&= \frac{1}{2} r^2 B_0 \sin^2\th.
\end{align*}

We now compute the electric and magnetic fields of this solution. Recalling that
\begin{equation}
E^i = F^{0i}, \quad 
B^i = (\star F)^{i0} = -\frac{1}{2}\wt{\e}^{0ijk} F_{jk},
\end{equation}
where $\wt{\e}_{0ijk} = \sqrt{|g|}\e_{0ijk}$, we find
\begin{align}
E^r &= \frac{Q}{r^2} - E_0\sin\th \cos\f \quad 
&B^r &= B_0 \cos\th \nonumber \\
E^\th &= -\frac{1}{r} E_0 \cos\th \cos\f \quad 
& B^\th &= -\frac{1}{r} B_0 \sin\th\nonumber \\
E^\f &= E_0 \frac{\sin\f}{r\sin\th}
\quad & B^\f &= 0.
\end{align}
We see that the $\th$ and $\f$ components are not affected by the black hole, and at first glance it appears that the $r$ components are not either. However, the coordinate basis vector $\bt{e}_r = \del_r$ in Schwarzschild spacetime time only becomes the standard spherical unit vector in the $r\to\infty$ limit. This can be made more clear by normalizing the basis vector. We find
\begin{equation}
\hbt{e}_r = N \bt{e}_r,
\end{equation}
so for the electric field
\begin{equation}
E^r \bt{e}_r = \left(\frac{Q}{r^2} 
- E_0 \sin\th \cos\f\right) N^{-1} \hbt{e}_r.
\end{equation}
This yields the expected electric field in flat space only in the $r\to\infty$ limit. The same is true for the magnetic field. 

We can compute the charge using Gauss's law. In the language of differential forms, we have
\begin{equation}
4\p Q_G = \int_V \star J = \int_V \text{d}(\star F) 
= \oint_{\del V} \star F.
\end{equation}
Taking $V$ to be a sphere of radius $R$, we find
\begin{align}
4\p Q_G 
&= \oint_{r=R} \left(\star F\right)_{\th\f} d\th \wedge d\f \nonumber\\
&= \int_{S^2} R^2\sin\th \, \del_r A_t \vert_{r=R} \, d\th\, d\f \nonumber\\
&= \int_{S^2} R^2\sin\th \, 
\left(\frac{Q}{R^2} 
- E_0 \sin\th \cos\f\right) \, d\th\, d\f \nonumber\\
&= Q \int_{S^2} \sin\th\, d\th\, d\f
- E_0 R^2 \int_{S^2} \sin^2\th \cos\f \, d\th\, d\f \nonumber\\
&= 4\p Q,
\end{align}
as we should.
We also compute the invariant 
\begin{align}
\bt{E} \cdot \bt{B} 
&= \frac{1}{4}F^{\m\n} (\star F)_{\m\n} \nonumber\\
&= \frac{Q B_0}{r^2}\cos\th 
- E_0 B_0 \frac{2M}{r}\cos\th \sin\th \cos\f.
\end{align}

\section{Initial Conditions}

We want to set the initial conditions in such a way that an observer at rest at a given point measures the velocity of the test charge at that point to be $\bt{v} = \b \hat{y}$. To this end, we build an orthonormal basis for the observer. If $e^\m_a = \left(\del_a\right)^\m$ denotes the $\m$th component of basis vector $a$, an orthonormal basis for such an observer is 
\begin{equation}
\hat{e}^\m_t 
= \frac{1}{\sqrt{1-\frac{2M}{r_0}}}e^\m_t, \quad
\hat{e}^\m_r = \sqrt{1-\frac{2M}{r_0}} e^\m_r, \quad
\hat{e}^\m_\th = \frac{1}{r_0} e^\m_\th, \quad
\hat{e}^\m_\f = \frac{1}{r_0\sin\th_0} e^\m_\f,
\end{equation}
which obeys $g_{\m\n} \hat{e}^\m_a \hat{e}^\m_b = \h_{ab}.$ The components of the four-velocity of the particle in the orthonormal basis should be $\g (1, \bt{v})$ with $\bt{v} = \b\hbt{y}$. Thus the initial conditions are
\begin{equation}
u_\m \hat{e}^\m_t = \frac{1}{\sqrt{1-\b^2}}, \quad
u_\m \hat{e}^\m_r = \frac{\b\sin\th_0\sin\f_0}{\sqrt{1-\b^2}}, \quad
u_\m \hat{e}^\m_\th = \frac{\b\cos\th_0\sin\f_0}{\sqrt{1-\b^2}}, \quad
u_\m \hat{e}^\m_\f = \frac{\b\cos\f_0}{\sqrt{1-\b^2}},\nonumber 
\end{equation}
which we can write as
\begin{equation}
\dot{t}(0)
= \frac{1}{\sqrt{\left(1-\b^2\right)
\left(1-\frac{2M}{r_0}\right)}}, 
\quad
\dot{r}(0)
= \frac{\b\sin\th_0\sin\f_0}{\sqrt{1-\b^2}}
\sqrt{1-\frac{2M}{r_0}},
\end{equation}
\begin{equation}
\dot{\th}(0) 
= \frac{\b\cos\th_0\sin\f_0}{r_0\sqrt{1-\b^2}}, 
\quad
\dot{\f}(0)
= \frac{\b\cos\f_0}{r_0\sin\th_0\sqrt{1-\b^2}}.
\nonumber
\end{equation}
These are the initial conditions used in all the simulations above.

\bibliographystyle{utphys}
\bibliography{refs}

\end{document}